\documentclass[fleqn,usenatbib]{mnras}

\usepackage{newtxtext,newtxmath}
\usepackage[T1]{fontenc}

\DeclareRobustCommand{\VAN}[3]{#2}
\let\VANthebibliography\thebibliography
\def\thebibliography{\DeclareRobustCommand{\VAN}[3]{##3}\VANthebibliography}

\usepackage{graphicx}	
\usepackage{amsmath}	

\usepackage{breakurl}

\title[Bar rotation rate and environmental density]{Bars in low-density environments rotate faster than bars in dense regions}

\author[N. Puczek et al.]{
Natalia Puczek,$^{1}$
Tobias G{\'e}ron,$^{1,2}$
Rebecca J. Smethurst$^{1}$
and Chris J. Lintott$^{1}$\thanks{E-mail: cjl@astro.ox.ac.uk (CJL)}
\\
$^{1}$Oxford Astrophysics, Department of Physics, University of Oxford, Denys Wilkinson Building, Keble Road, Oxford OX1 3RH, UK\\
$^{2}$Dunlap Institute for Astronomy \& Astrophysics, University of Toronto, 50 St. George Street, Toronto ON M5S 3H4, Canada
}

\date{Accepted XXX. Received YYY; in original form ZZZ}

\pubyear{\the\year{}}

\begin{document}
\label{firstpage}
\pagerange{\pageref{firstpage}--\pageref{lastpage}}
\maketitle

\begin{abstract}
Does the environment of a galaxy directly influence the kinematics of its bar? We present observational evidence that bars in high-density environments exhibit significantly slower rotation rates than bars in low-density
environments. Galactic bars are central, extended structures composed of stars, dust and gas, present in approximately 30 to 70 per cent of luminous spiral galaxies in the local Universe. Recent simulation studies have suggested that the environment can influence the bar rotation rate, $\mathcal{R}$, which is used to classify bars as either fast ($1\leq\mathcal{R}\leq1.4$) or slow ($\mathcal{R}>1.4$). We use estimates of $\mathcal{R}$ obtained with the Tremaine--Weinberg method applied to Integral Field Unit spectroscopy from MaNGA and CALIFA. After cross-matching these with the projected neighbour density, $\log\Sigma$, we retain 286 galaxies. The analysis reveals that bars in high-density environments are significantly slower (median $\mathcal{R} = 1.65^{+0.13}_{-0.11}$) compared to bars in low-density environments (median $\mathcal{R} =1.39^{+0.09}_{-0.08}$); Anderson--Darling \textit{p}-value of $p_{\mathrm{AD}}= 0.002$ ($3.1\,\sigma$). This study marks the first empirical test of the hypothesis that fast bars are formed by global instabilities in isolated galaxies, while slow bars are triggered by tidal interactions in dense environments, in agreement with predictions from numerous \textit{N}-body simulations. Future studies would benefit from a larger sample of galaxies with reliable Integral Field Unit data, required to measure bar rotation rates. Specifically, more data are necessary to study the environmental influence on bar formation within dense settings (i.e. groups, clusters and filaments).
\end{abstract}


\section{Introduction}

A bar in a galaxy is a central, extended, long-lived structure composed of stars, dust and gas, present in approximately 30 to 70 per cent of luminous spiral galaxies in the local Universe, with the bar fraction depending on factors such as bar strength, wave-band in which it is observed, galaxy colour and mass 
(\citealt{2000AJ....119..536E,2008ApJ...675.1141S,2011MNRAS.411.2026M,2018MNRAS.474.5372E,2021MNRAS.507.4389G}). Although \textit{JWST} has now detected bars beyond ${z\sim 3}$ (a look-back time of $11.5\,\mathrm{Gyr}$; \citealt{2023Natur.623..499C, 2023ApJ...958...36S,2025MNRAS.537.1163A,2025ApJ...987...74G}), simulations mark the epoch of bar formation to be around $z \sim 0.7-1$, coinciding with the galaxies’ transition to being dynamically cool and disc-dominated and a decline in major mergers (\citealt*{2012ApJ...757...60K}; \citealt{2014MNRAS.438.2882M}). In this secular epoch, slow processes become more dominant in galaxy evolution, providing a stable environment for bars to form and persist over long timescales.

Galactic bars are thought to play a critical role in galaxy evolution, funnelling gas to the centre of the galaxy, triggering starbursts and facilitating quenching (\citealt{2012MNRAS.424.2180M,2013ApJ...779..162C, 2018MNRAS.473.4731K, 2021MNRAS.507.4389G}). Bars allow mass and angular momentum redistribution, stablizing stellar orbits and influencing subsequent evolution of disc galaxies \citep{1973ApJ...186..467O,2009ASPC..419...31C,2012ApJ...751...44S}. 
Therefore, studying bars is pivotal to building a coherent description of galaxy dynamics.

Both galaxy stellar mass and dark matter halo mass have been suggested as important factors in bar formation and evolution (\citealt{2018MNRAS.477.1451F,2020MNRAS.491.2547R}; \citealt*{2022MNRAS.509.1262K}; \citealt{2023ApJ...947...80B}). Although some studies argue that bar formation and evolution are predominantly influenced by the internal processes of the host galaxy (\citealt*{2021MNRAS.501..994S}; \citealt{2023A&A...679A...5A}), the dynamical impact of events such as flybys, minor mergers and satellite interactions on the formation and evolution of barred galaxies has been extensively studied (\citealt{2016MNRAS.463.2210K, 2017A&A...604A..75M,2020A&A...641A..77C,2021MNRAS.502.3085G}). It has long been established that bar formation can be triggered by either tidal interactions (\citealt{1987MNRAS.228..635N}; \citealt*{2014ApJ...790L..33L}; \citealt{2016ApJ...826..227L}; \citealt*{2017ApJ...842...56G}; \citealt{2018ApJ...857....6L}) or by internal instabilities  in cold isolated discs (\citealt{1971ApJ...168..343H, 1973ApJ...186..467O}; \citealt*{2013MNRAS.429.1949A}; \citealt{2014RvMP...86....1S}). Simulations further show that tidal interactions can trigger bar formation in galaxies, which would have not otherwise been able to form a bar in isolation (\citealt{1998ApJ...499..149M,2014ApJ...790L..33L,2017MNRAS.464.1502M}). While the internal dynamics and morphologies of hosts are an important factor in the study of bars, bar formation and evolution are also strongly linked to the environment of the galaxy. 

However, observational evidence for a relationship between environmental density and bar pattern speed has been lacking. Bar pattern speed, $\Omega _{\mathrm{bar}}$, is a fundamental parameter of a bar, as it measures the angular frequency of the bar's rotation about the galactic centre. Given the variation in galaxy sizes, bar pattern speed is normally parametrised by the bar rotation rate, defined as the dimensionless ratio of the bar’s corotation radius, $R_{\mathrm{cr}}$, and its semi-major axis, $R_{\mathrm{bar}}$ \citep{2019AA...632A..51C}, that is 
\begin{equation}
    \mathcal{R} = \frac{R_{\mathrm{cr}}}{R_{\mathrm{bar}}}.
\end{equation}
The corotation radius is the radius at which the bar and the stars in the disc rotate at the same speed. The bar rotation rate parameter is used to classify bars as either slow ($\mathcal{R}>1.4$) or fast ($1\leq\mathcal{R}\leq1.4$), with the median value of $\mathcal{R}$ being around 1.66 \citep{thesis}. Slow bars fall short of $R_{\mathrm{cr}}$, while fast bars extend out to the corotation radius and their pattern speeds are close to the maximum rotation speed allowed at a given bar length. Despite ultrafast bars ($\mathcal{R}<1$) being unphysical \citep{1980A&A....81..198C,1992MNRAS.259..345A}, they are repeatedly measured observationally (\citealt*{2008MNRAS.388.1803R}; \citealt{2017ApJ...835..279F,2019MNRAS.482.1733G}). This is likely a measurement artefact, as the ultrafast bar phenomenon essentially disappears when bar radius is measured carefully (\citealt{2021MNRAS.503.2833R,2021A&A...649A..30C}). In this paper, we refer to all bars with $\mathcal{R}\leq1.4$ as fast.

\textit{N}-body simulations indicate that bars triggered by instabilities within the discs of isolated galaxies are typically fast, whereas bars formed due to tidal interactions are slow (\citealt{1998ApJ...499..149M,2004MNRAS.347..220B,2014MNRAS.445.1339L,2016ApJ...826..227L,2017MNRAS.464.1502M,2017ApJ...842...56G,2018ApJ...857....6L}). Simulations show that bar evolution in isolated galaxies proceeds through three distinct phases. Once the bar emerges, it continues to grow in the vertical direction for the first ${\sim\,2\,\mathrm{Gyr}}$, after which it buckles (bends out of the galactic disc forming a boxy/peanut bulge) becoming shorter and weaker for the next $\sim\,1\,\mathrm{Gyr}$ (\citealt{1981A&A....96..164C, 1991Natur.352..411R,2004ApJ...613L..29M}). Eventually it enters a phase during which it slows down and gradually grows in length and strength over several Gyr (\citealt{1991Natur.352..411R}; \citealt*{2006ApJ...637..214M}; \citealt{2013seg..book..305A,2014RvMP...86....1S}). $\mathcal{R}$ fluctuates across all three stages, but in general, $\mathcal{R} \lesssim 1.4$ for most of the time (\citealt{1998ApJ...499..149M,2004MNRAS.347..220B,2017MNRAS.464.1502M}). Simulations of bars influenced or triggered by tidal interactions (i.e. tidal bars) indicate that tidal bars initially undergo weak buckling, after which their rotation rates, $\mathcal{R}$, decrease gradually for $\sim4\, \mathrm{Gyr}$ (\citealt{1998ApJ...499..149M,2017MNRAS.464.1502M,2017ApJ...842...56G,2018ApJ...857....6L}), such that the bars eventually enter the fast regime, with $\mathcal{R} \sim 1.4$ (\citealt{2017MNRAS.464.1502M}). None the less, for most of the time, tidal bars are very slow, with $\mathcal{R}\sim 2\,{-}\,3$  (\citealt{1998ApJ...499..149M,2004MNRAS.347..220B,2014MNRAS.445.1339L, 2016ApJ...826..227L, 2017ApJ...842...56G, 2018ApJ...857....6L}). \citet{1998ApJ...499..149M} suggest angular momentum transfer to the perturber as a possible explanation for this slow rotation.

Therefore, $\mathcal{R}$ could be an observational parameter which distinguishes between tidal bars and bars triggered by internal instabilities. \citet{2016ApJ...826..227L} simulated bar formation on different orbits in a cluster, demonstrating that tidal interactions can trigger or influence bar formation in cluster cores but not in cluster outskirts. Similarly, tidal features in galaxies have been shown to be more frequent in groups and clusters compared to isolated galaxies (\citealt{1988A&A...189...42F}). Therefore, we expect that a higher environmental density would lead to a higher fraction of tidal bars. We hypothesise that measurements of bar rotation rate and environmental density should reveal a positive correlation between the two variables. This study is the first observational test of this hypothesis.

To determine whether the simulation results are confirmed by observational data, we use estimates of the rotation rate, $\mathcal{R}$, obtained for 334 galaxies in the local Universe using the Tremaine--Weinberg method applied to Integral Field Unit (IFU) spectroscopy from Mapping Nearby Galaxies at Apache Point Observatory (MaNGA; \citealt{2015ApJ...798....7B}) and Calar Alto Legacy Integral Field Area (CALIFA; \citealt{2012A&A...538A...8S}) surveys, across six studies (\citealt{2015AA...576A.102A,2019MNRAS.482.1733G,2019AA...632A..51C,2020MNRAS.491.3655G,2022MNRAS.517.5660G,2023MNRAS.521.1775G}). We cross-match these with projected neighbour densities, $\log _{10} (\Sigma / \textrm{Mpc}^{-2})$ (i.e. $\log~\Sigma$), from \citet{2006MNRAS.373..469B} using a 10 arcsec search radius, retaining 286 galaxies. To test whether galaxy stellar mass could be responsible for variations in $\mathcal{R}$, we additionally cross-match the $\log~\Sigma$ sample with stellar masses, $\log_{10} \left( M_\star / \mathrm{M}_\odot\right)$ (i.e. $\log M_\star$), from the MPA-JHU  collaboration (Max-Planck Institute for Astrophysics \& John Hopkins University; \citealt{2003MNRAS.341...33K}), retaining 275 galaxies. This marks the first empirical test of the assumption that fast bars are formed by global instabilities in the galactic discs of isolated galaxies, while slow bars are triggered by tidal interactions in dense environments.

This paper proceeds as follows: In Section~\ref{data} we provide a detailed description of our data sources, galaxy sample and analysis methods. In Section~\ref{results} we present the measured relationships between $\mathcal{R}$, $\log~\Sigma$ and $\log M_\star$. We discuss the results in Section~\ref{diss}. Finally, we summarise our findings in Section~\ref{con}.

\section{DATA AND METHODS} \label{data} 
\subsection{Bar rotation rate $\mathcal{R}$ and Tremaine--Weinberg method}

We obtain the dimensionless $\mathcal{R}$ measurements for 210 galaxies from \citet{2023MNRAS.521.1775G}, who applied the model-independent kinematic Tremaine--Weinberg (TW) method based on the work of \citet{1984ApJ...282L...5T} to line-of-sight stellar velocity and stellar flux measurements. The TW method measures the bar pattern speed, $\Omega _ \mathrm{bar}$, by calculating the velocity component misalignment with the major axis of the galaxy. To determine the corotation radius, \citet{2023MNRAS.521.1775G} combined their measurements of $\Omega _ \mathrm{bar}$ with galaxy rotation curves derived from stellar velocity data. For each galaxy, the authors multiplied $\Omega _ \mathrm{bar}$ by a range of radii, obtaining a measure of the bar's speed as a function of radius. The radius at which this curve intersected the galaxy rotation curve was the corotation radius. \citet{2023MNRAS.521.1775G} subsequently combined their measurements of $R_{\mathrm{cr}}$ with manual bar length measurements to calculate  $\mathcal{R}$. The authors obtained stellar velocity and stellar flux data from MaNGA IFU in the seventeenth data release of the Sloan Digital Sky Survey (SDSS; \citealt{2022ApJS..259...35A}). Additionally, they made use of Galaxy Zoo DESI \citep{2023MNRAS.526.4768W} to identify barred galaxies in the IFU survey.

\citet{2023MNRAS.521.1775G} provide the largest sample of kinematically classified bars to date. However, given the rigorous constraints on data necessary for the TW method, this is still only a small fraction of bars observed by MaNGA. Specifically, the method can only be applied to galaxies with intermediate inclinations ($20 ^{\circ}< i < 70 ^{\circ}$), since edge-on galaxies do not have accurate spatial measurements, while face-on galaxies lack accurate stellar velocity measurements. Additionally, the bar may not be aligned with the axes of the galaxy, presenting an additional constraint (\citealt{1984ApJ...282L...5T,2023MNRAS.521.1775G}). As a result, the authors were able to retain only 2~per~cent of MaNGA barred galaxies identified by Galaxy Zoo DESI. Therefore, we supplement the galaxy sample with $\mathcal{R}$ measurements from five other studies which also applied the TW method, retaining only the most recent $\mathcal{R}$ estimate in cases of duplicates, in the same order as they appear in Table~\ref{tab:tab}. This results in a sample containing 334 galaxies with measured $\mathcal{R}$. Our sample covers the local Universe, with a redshift range $0.017 < z <0.078$.

\begin{table}
	\centering
	\caption{Summary of the bar rotation studies used in our study. The table outlines the IFU spectroscopy catalogue used in the studies and the number of galaxies added to our bar rotation rate, projected neighbour density, and galaxy stellar mass samples.}
	\label{tab:example_table}
	\begin{tabular}{lp{1.24cm}ccc} 
		\hline
		Study & Catalogue & $\mathcal{R} $& $ \log~\Sigma$ & $ \log M_\star$\\
		\hline
		\citet{2023MNRAS.521.1775G} & MaNGA & 210 & 190 & 185\\
		\citet{2022MNRAS.517.5660G} & MaNGA & 56 & 51 & 46\\
		\citet{2020MNRAS.491.3655G} & MaNGA \& CALIFA & 6 & 3 & 3\\
  		\citet{2019AA...632A..51C} & CALIFA & 16 & 9 & 9\\
            \citet{2019MNRAS.482.1733G} & MaNGA & 34 & 28 & 27\\
		\citet{2015AA...576A.102A} & CALIFA & 12 & 5 & 5\\
  \hline \hline
  		 & \textbf{TOTAL} & 334 & 286 & 275\\
	\end{tabular}
 \label{tab:tab}
\end{table}

\subsection{Local environmental density $\log\,\Sigma$} \label{log_Sigma}

We use the dimensionless projected neighbour density, $\log_{10}~(\Sigma / \textrm{Mpc}^{-2})$, from \citet{2006MNRAS.373..469B}, as a measure of environmental density (measured using spectroscopy from the seventh release of SDSS; \citealt{2009ApJS..182..543A}). The projected neighbour density is available for 286 galaxies from our original sample. $\log~\Sigma$ is derived by averaging $\log_{10}~(\Sigma _N / \textrm{Mpc}^{-2})$  for $N = 4$ and~5, where 
\begin{equation}
    \Sigma_{N} = \frac{N}{\pi d_N^2},
\end{equation}
and $d_N \left[ \textrm{Mpc}\right]$ is the projected comoving distance to the $N^{\textrm{th}}$ nearest neighbour. A larger (smaller) value of $\log~\Sigma$ corresponds to a denser (less dense) environment. The galaxies in our sample have local densities in the range $-1.46<\log\,\Sigma<1.92$. The median measurement uncertainty in our sample is 0.05 dex. 

As our sample size is small, with galaxies in the highest-density environments underrepresented, we set cut-offs for the low, intermediate and high-density environments at the 33.3rd and 66.7th percentiles of $\log~\Sigma$ ($-0.46$ and $0.13$, respectively). This approach produces roughly equal-sized density bins. Our low-density bin has a higher cut-off than the $\log~\Sigma < -0.8$ cut-off sometimes used to distinguish voids (\citealt{2006MNRAS.373..469B}; \citealt*{2007MNRAS.382..801M}). However, it aligns with how isolated galaxies are defined by \citet{2010A&A...509A..40I}, who use the $-0.8 < \log~\Sigma < -0.4$ range. \citet{2009MNRAS.393.1324B} apply an even more generous cut-off of $\log~\Sigma < 0$ for low-density environments. Our high-density bin ($\log~\Sigma > 0.13$) falls below typical cluster ranges. \citet{2007MNRAS.382..801M} use $\log~\Sigma > 0.8$ for inner parts of galaxy clusters and \citet{2006MNRAS.373..469B} use $\log~\Sigma > 0.8$ for cluster-like environments. It also falls below the lower limit for group environments ($0.4 < \log~\Sigma < 0.8$) set by \citet{2010A&A...509A..40I}. Therefore, our low, intermediate and high-density labels refer to relative sample density, with the low-density bin predominantly containing isolated galaxies in void-like regions, and the high-density bin containing a range of dense environments (including groups and clusters). 

To test the robustness of our results against the choice of density cut-offs, we repeat the analysis using density cuts shifted to the nearest $0.2$ dex relative to the 33.3rd and 66.7th percentiles of $\log~\Sigma$ ($-0.4$ and $0.2$). We also explore other subdivisions of $\log~\Sigma$ (outlined in Table~\ref{tab:tab2}). We verify that the galaxy stellar mass distributions of the low and high-density subsamples are not statistically distinguishable for each $\log~\Sigma$ subdivision (Anderson--Darling \textit{p}-value $<1.75\,\sigma$). This justifies our decision not to apply mass matching, which would further reduce sample sizes. Such an approach ensures that our results are robust to reasonable variations in sample definition and are unlikely to be driven by sample-selection biases.

\subsection{Galaxy stellar mass $ \log M_\star $}

We use the dimensionless logarithmic median total stellar masses, $\log _{10}\left( M_\star / \mathrm{M}_\odot\right)$, from the MPA-JHU collaboration (\citealt{2003MNRAS.341...33K}). The authors derived the masses through fits to the seventh release of SDSS photometry \citep{2009ApJS..182..543A}. Stellar masses are available for 275 galaxies from the $\log~\Sigma$ sample described in the previous paragraph. The galaxies in the $\log\,\Sigma$ sample have stellar masses in the range $9.38<\log M_\star<12.05$. The median measurement uncertainty in our sample is $0.09$ dex. 

In literature, there are no agreed-upon thresholds for galaxy stellar mass which we could use to separate our sample into low, intermediate and high-mass bins. For example, \citet{2020MNRAS.491.3318S} use a $10^{11}\,\mathrm{M}_\odot$ threshold to define high-mass galaxies, while  \citet{2025A&A...695A..84P} use $10^{10.5}\,\mathrm{M}_\odot$ as a cut-off for low-mass and high-mass galaxies. \citet{2021MNRAS.507.4389G} refer to the $M_\star<10^{10}\,\mathrm{M}_\odot$ range as low. A reasonable threshold is $M_\star\,\sim 10^{10.2}\,\mathrm{M}_\odot~-~10^{10.3}\, \mathrm{M}_\odot $, which marks a reversal of bar fraction trends in the local Universe (\citealt{2010ApJ...714L.260N,2012MNRAS.423.1485S,2025MNRAS.542..151M}). However, because of the small sample size in our study, we choose to set the cut-offs at the 33.3rd and 66.7th percentiles of $ \log M_\star $, producing roughly equal-sized galaxy stellar mass bins. This corresponds to galaxy stellar masses of $10^{10.51}\,\mathrm{M}_\odot$ and $10^{10.85}\,\mathrm{M}_\odot$, respectively. Therefore, our low-mass bin is consistent with the one used in \citet{2025A&A...695A..84P}, whereas the high-mass cut-off lies between the ones used by \citet{2020MNRAS.491.3318S} and \citet{2025A&A...695A..84P}.

\subsection{Statistical analysis}
To assess the significance of differences between subgroups, we perform the Anderson--Darling test. The null hypothesis is that the values in the low and high-density, or equivalently the low and high-mass subgroups, are drawn from the same population. We require $3\,\sigma$ for statistical significance. We calculate medians of subsamples and their respective uncertainties via bootstrap resampling 5000 times with replacement. We verify that increasing the number of resamples beyond 5000 does not change the inferred medians and confidence intervals. We compute the median for each iteration, and obtain the final median and $1\,\sigma$ uncertainties from the 50th, 16th, and 84th percentiles of the bootstrap distribution. 
\renewcommand{\arraystretch}{1.3}
\begin{table*}
\centering
\caption{Median values of $\mathcal{R}$ in low, intermediate, and high-density environments ($\mathcal{R}_{\rm low}$, $\mathcal{R}_{\rm mid}$ and $\mathcal{R}_{\rm high}$) with $1\,\sigma$ uncertainties for different subdivisions of $\log~\Sigma$ . $N_{\rm low}$, $N_{\rm mid}$ and $N_{\rm high}$ are the corresponding sizes of the subgroups. $p_{\rm AD}$ is the Anderson--Darling \textit{p}-value between the low and high-density subgroups. (1): 33.3rd and 66.7th percentiles, (2): 33.3rd and 66.7th percentiles shifted to the nearest $0.2$ dex, (3): 50th percentile, (4): isolated; groups (\citealt{2010A&A...509A..40I}), (5): voids; clusters (\citealt{2006MNRAS.373..469B,2007MNRAS.382..801M}), (6): low-density environments with $\log\,\Sigma < 0$ (\citealt{2009MNRAS.393.1324B}).}
\label{tab:tab2}
\begin{tabular}{lccccccc}
\hline
$\log \Sigma$ cut-offs &
$\mathcal{R}_{\rm low}$ &
$N_{\rm low}$ &
$\mathcal{R}_{\rm mid}$ &
$N_{\rm mid}$ &
$\mathcal{R}_{\rm high}$ &
$N_{\rm high}$ &
$p_{\rm AD}$ \\
\hline
(1) $-0.46;\, 0.13$ &
$1.39^{+0.09}_{-0.08}$ & 95 &
$1.52^{+0.11}_{-0.10}$ & 96 &
$1.65^{+0.13}_{-0.11}$ & 95 &
$0.0019\,(3.1\sigma)$ \\

(2) $-0.4;\, 0.2$ &
$1.40^{+0.09}_{-0.08}$ & 106 &
$1.53^{+0.11}_{-0.10}$ & 97 &
$1.66^{+0.14}_{-0.12}$ & 83 &
$0.0028\,(3.0\sigma)$ \\

(3) $-0.15$ &
$1.40^{+0.08}_{-0.07}$ & 143 &
-- & -- &
$1.63^{+0.10}_{-0.09}$ & 143 &
$0.0032\,(2.9\sigma)$ \\

(4) $-0.4;\, 0.4$&
$1.40^{+0.09}_{-0.08}$ & 106 &
$1.54^{+0.09}_{-0.09}$ & 124 &
$1.71^{+0.20}_{-0.16}$ & 56 &
$0.0033\,(2.9\sigma)$ \\

(5) $-0.8;\, 0.8$&
$1.40^{+0.15}_{-0.12}$ & 45 &
$1.52^{+0.07}_{-0.06}$ & 221 &
$1.68^{+0.25}_{-0.22}$ & 20 &
$0.065\,(1.8\sigma)$ \\

(6) $0$&
$1.45^{+0.07}_{-0.07}$ & 175 &
-- & -- &
$1.62^{+0.11}_{-0.10}$ & 111 &
$0.015\,(2.4\sigma)$ \\
\hline
\end{tabular}
\end{table*}

To account for large asymmetric measurement uncertainties in $\mathcal{R}$ (median individual $1\,\sigma$ measurement uncertainties of $-0.37$ and $+0.59$), we explicitly propagate them. In each of the 5000 resamples, we perturb the measured values of $\mathcal{R}$ using a distribution constructed from two half-normal distributions joined at the measured value of $\mathcal{R}$. Such modeling of posterior distributions of $\mathcal{R}$ ensures that we draw values above and below $\mathcal{R}$ with equal probability. In other words, this ensures that the measured values of $\mathcal{R}$ coincide with the medians of the modeled posterior. As such, we sample the downward (upward) perturbations from half-normal distributions with parametrisations set to the reported negative (positive) measurement uncertainties. By construction, this method ensures that the medians and the 16th and 84th percentiles of the modeled posterior distributions coincide with the measured values of $\mathcal{R}$ and their corresponding lower and upper bounds.

To add a simple measure of the strength of the relationship between $\mathcal{R}$ and $\log\,\Sigma$, we also calculate the Pearson correlation coefficient, $r$, along with its \textit{p}-value, requiring $3\,\sigma$ for statistical significance. While $r$ does not account for measurement uncertainties or intrinsic scatter, it is an easily interpretable metric. As all bars go through multiple phases of evolution, with rotation rates fluctuating across them, we expect substantial scatter in each density bin rather than a tight, one-to-one relation between $\mathcal{R}$ and $\log\,\Sigma$. To quantify the relationship between the two variables, we use the bivariate correlated errors and intrinsic scatter regression, BCES y|x (\citealt{1996ApJ...470..706A}). This method accounts for both measurement uncertainties and intrinsic scatter and is a direct extension of the ordinary least-squares approach.

In Section~\ref{results}, we investigate the dependence of the bar rotation rate on projected neighbour density and galaxy stellar mass by examining the relationship between $\mathcal{R}$ and $\log~\Sigma$, as well as comparing the distributions of $\mathcal{R}$ across low and high-density environments and between low and high-mass galaxies. We discuss the results of these tests in Section~\ref{diss}.

\section{Results} \label{results}

\begin{figure}
	\includegraphics[width=\columnwidth]{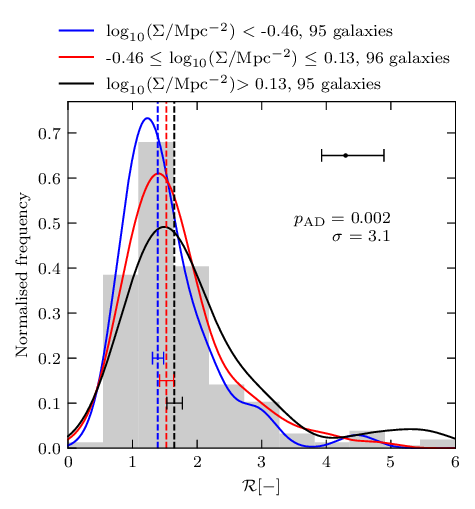}
    \caption{Distributions of the bar rotation rate, $\mathcal{R}$,
    for 286 galaxies. The error bar in the top-right corner represents the median  measurement uncertainty in $\mathcal{R}$ for individual galaxies. The distributions for the low, intermediate and high-density samples fitted with Gaussian kernel estimates are blue, red and black, respectively. The dashed vertical lines indicate the bootstrapped median values of $\mathcal{R}$ in the samples. The three error bars on the dashed vertical lines represent the bootstrapped asymmetric $1\,\sigma$ uncertainties on the values of median $\mathcal{R}$ in each environment. The Anderson--Darling \textit{p}-value with the null hypothesis being that the $\mathcal{R}$ values in the low and high-density samples are drawn from the same population is $p_{\mathrm{AD}} = 0.002$, a statistically significant result ($3.1\,\sigma$). This means that galaxies in low-density environments preferentially host faster bars (with lower $\mathcal{R}$) than galaxies in high-density environments, despite a significant overlap between the two distributions.}
    \label{fig:r-sigm}
\end{figure}

\subsection{Local environmental density $\log\,\Sigma$}
The distributions of $\mathcal{R}$ in the low and high-density subgroups, set by the cut-offs at the 33.3rd and 66.7th percentiles of $\log\,\Sigma$, are shown in Fig.~\ref{fig:r-sigm}. The median values of $\mathcal{R}$ in the low, intermediate and high-density environments follow a clear positive trend, with $\mathcal{R}~=~1.39^{+0.09}_{-0.08}$, $\mathcal{R}~=~1.52^{+0.11}_{-0.10}$ and $\mathcal{R}~=~1.65^{+0.13}_{-0.11}$, respectively. To assess the significance of the difference between the low-density and high-density environments, we perform the Anderson--Darling test. The null hypothesis is that the $\mathcal{R}$ values in the low and high-density subgroups are drawn from the same population. Both subgroups contain 95 galaxies. We obtain a \textit{p}-value of $p_{\mathrm{AD}}=0.002$, which corresponds to a 3.1$\,\sigma$ result. Since we require at least 3$\,\sigma$ for statistical significance, the Anderson--Darling test indicates that bars hosted in galaxies in low-density environments have significantly lower values of $\mathcal{R}$ (i.e. tend to be faster) than galaxies in high-density environments, despite a big overlap between the two distributions (see Fig.~\ref{fig:r-sigm}).

We repeat the analysis using other subdivisions of $\log~\Sigma$ to test the robustness of our results (see Table~\ref{tab:tab2}). Shifting the fiducial local density cuts to the nearest $0.2$ dex (cut 2) preserves the values of median $\mathcal{R}$ to within $1\%$, maintains the positive trend across the low, intermediate and high-density subgroups, and yields a significant Anderson--Darling test ($p_{\mathrm{AD}}=0.003$, $3.0\,\sigma$). Using the 50th percentile of $\log\,\Sigma$ (cut 3) or the $\log\,\Sigma=-0.4,0.4$ cuts (cut 4; \citealt{2010A&A...509A..40I}) results in slightly lower Anderson--Darling significances ($2.9\,\sigma$), likely due to reduced contrast between subsamples and lower high-density sample size, respectively. None the less, the positive trend in $\mathcal{R}$ and $\log\,\Sigma$ is maintained. The extreme $\log\,\Sigma=-0.8, 0.8$ density cuts (cut 5; \citealt{2006MNRAS.373..469B,2007MNRAS.382..801M}) result in very small sample sizes ($N_{\rm low} = 45$ and $N_{\rm high} = 20$) and a nonsignificant Anderson--Darling test ($p_{\mathrm{AD}}=0.065$, $1.8\,\sigma$), while using $\log\,\Sigma = 0$ as a threshold for low/high local densities (cut 6; \citealt{2009MNRAS.393.1324B}) results in moderate Anderson--Darling test significance ($p_{\mathrm{AD}}=0.015$, $2.4\,\sigma$). In all cases, median $\mathcal{R}$ consistently show a positive trend with $\log\,\Sigma$, demonstrating that our results are robust to reasonable variations in the choice of $\log\,\Sigma$ cut-offs. 

\begin{figure}
	\includegraphics[width=\columnwidth]{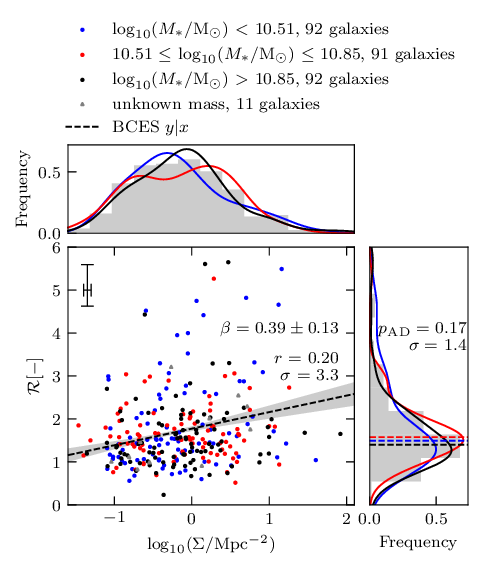}
    \caption{Scatter plot of the bar rotation rate, $\mathcal{R}$, against projected neighbour density, $\log~\Sigma$, for 286 galaxies. The error bars represent the median measurement errors in $\mathcal{R}$ and $\log~\Sigma$ for individual galaxies. Low, intermediate and high-mass galaxies are coloured blue, red and black, respectively (the same colouring has been used for the histograms). 11 galaxies with unknown mass are coloured grey. The black dashed line shows a robust linear regression obtained with the BCES $y|x$ method with the grey shaded region representing $1\,\sigma$ confidence. The best-fitting slope to the data obtained with this method is $\beta = 0.39 \pm 0.13$. The Pearson correlation coefficient between $\mathcal{R}$ and $\log~\Sigma$ is $r~=~0.20$ with $3.3\,\sigma$ significance, indicating a statistically significant weakly positive correlation. The histograms show low, intermediate and high-mass galaxy samples fitted with Gaussian kernel estimates. The Anderson--Darling \textit{p}-value with the null hypothesis being that the $\mathcal{R}$ values in the low and high-mass samples are drawn from the same population is $p_{\mathrm{AD}} = 0.17$ ($1.4\,\sigma$), so the distributions are not statistically significantly different.}
    \label{fig:r-sigm-m}
\end{figure}

Fig.~\ref{fig:r-sigm-m} shows a scatter plot between $\mathcal{R}$ and $\log~\Sigma$ along with histograms depicting the distributions of both variables. The error bars represent the median measurement errors in $\mathcal{R}$ and $\log~\Sigma$ for individual galaxies. We see a weak positive trend between $\mathcal{R}$ and $\log~\Sigma$. A robust linear regression accounting for intrinsic scatter, obtained with the BCES $y|x$ method (bivariate correlated errors and intrinsic scatter; \citealt{1996ApJ...470..706A}), is depicted with the black dashed line. The best-fitting slope between $\mathcal{R}$ and $\log~\Sigma$ is $\beta = 0.39 \pm 0.13$ ($1\,\sigma$ uncertainty). To better capture the trend in the data, we calculate the Pearson correlation coefficient between $\mathcal{R}$ and $\log~\Sigma$ which yields a value of $r=0.20$ with 3.3$\,\sigma$ significance, confirming a statistically significant weakly positive correlation.

\subsection{Ultrafast bars $\mathcal{R} < 1$}
The fraction of bars with $\mathcal{R} < 1$ in the subgroups with density cut-offs at the 33.3rd and 66.7th percentiles of $\log~\Sigma$ decreases with increasing environmental density ($19$, $16$, and $11$ per cent from low to high local density), reinforcing the observed trend that bars in denser environments tend to have higher rotation rates. Excluding these bars reduces the significance of the Anderson--Darling test between $\mathcal{R}$ distributions in low and high-density environments from $3.1\,\sigma$ ($p_{\mathrm{AD}}=0.002$) to $2.7\,\sigma$ ($p_{\mathrm{AD}}=0.007$). Although this falls slightly below the $3\,\sigma$ required for statistical significance, the trend across the low, intermediate, and high-density subgroups remains consistent, with median $\mathcal{R}$ of $1.54^{+0.10}_{-0.10}$, $1.70^{+0.12}_{-0.11}$ and $1.78^{+0.15}_{-0.13}$, respectively. If we adopt a less strict definition of ultrafast bars, defined as bars which have a $1\,\sigma$ upper limit below $\mathcal{R} = 1$ (\citealt{2019MNRAS.482.1733G,2023MNRAS.521.1775G}), the ultrafast bar fractions in the low, intermediate and high-density environments drop to 4, 3 and 2 per cent, respectively. The median $\mathcal{R}$ in the environments become $\mathcal{R} = 1.43^{+0.09}_{-0.08}$, $\mathcal{R} = 1.56^{+0.12}_{-0.10}$ and $\mathcal{R} = 1.67^{+0.13}_{-0.11}$, and the Anderson--Darling test gives a statistically significant result, $p_{\mathrm{AD}}=0.003$ ($3.0\,\sigma$). While ultrafast bars may be measurement artefacts, we discuss in Section \ref{uf} why we choose not to exclude them from our analysis.

\subsection{Galaxy stellar mass $ \log M_\star $}
We next examine whether bar rotation rate depends on galaxy stellar mass. The data in the scatter plot and the histograms in Fig.~\ref{fig:r-sigm-m} are coloured based on the 33.3rd and 66.7th percentiles of $ \log M_\star $. The median values of $\mathcal{R}$ in the low, intermediate and high-mass subgroups are $\mathcal{R}~=~1.49_{-0.07}^{+0.07}$, $\mathcal{R}~=~1.58_{-0.06}^{+0.06}$ and $\mathcal{R}~=~1.40_{-0.07}^{+0.08}$, respectively. To assess the significance of the difference between the low-mass and high-mass environments, we perform the Anderson--Darling test. The null hypothesis is that the $\mathcal{R}$ values in the low and high-mass subgroups are drawn from the same population. We obtain a \textit{p}-value of $p_{\mathrm{AD}}~=~0.17$, which corresponds to $1.4\,\sigma$, indicating that the distributions do not differ significantly.

The above results suggest that the dynamics of bars depend more strongly on environmental density than on galaxy stellar mass, consistent with theoretical expectations and predictions from simulations.

\section{DISCUSSION} \label{diss}

Studies by \citet{1998ApJ...499..149M}, \citet{2004MNRAS.347..220B}, \citet{2014MNRAS.445.1339L,2016ApJ...826..227L},
\citet{2017MNRAS.464.1502M}, \citet{2017ApJ...842...56G} and \citet{2018ApJ...857....6L} suggest that tidal bars tend to rotate slower with respect to the stellar discs than bars formed by internal disc instabilities, possibly due to angular momentum transfer to the perturber during the interaction, as proposed by \citet{1998ApJ...499..149M}. Since tidal bar formation has been shown to be possible in cluster cores but not in cluster outskirts \citep{2016ApJ...826..227L}, we hypothesised that a higher environmental density would lead to a higher fraction of tidally triggered bars. Hence, if tidally triggered bars are slower, measurements of $\mathcal{R}$ and $\log~\Sigma$ should reveal a positive correlation between the two variables.  

\subsection{Local environmental density $\log\,\Sigma$}
To test this prediction with observational data, our approach was to compare the bar rotation rates of 286 galaxies across different environmental densities (see Fig.~\ref{fig:r-sigm}). We verified that the galaxy stellar mass distributions of the low and high-density subsamples are not statistically distinguishable for each $\log~\Sigma$ subdivision, justifying our decision not to apply mass matching. We have found that bars in high-density environments are significantly slower than bars in low-density environments ($p_{\mathrm{AD}}= 0.002$, $3.1\,\sigma$). Specifically, in dense environments ($\log\,\Sigma > 0.13$), we have measured the median $\mathcal{R} = 1.65^{+0.13}_{-0.11}$, whereas in low-density environments ($\log\,\Sigma < -0.46$), we have measured it to be $\mathcal{R} =1.39^{+0.09}_{-0.08}$. However, please note the overlap between the distributions in Fig.~\ref{fig:r-sigm} and the large median individual measurement uncertainty in $\mathcal{R}$ ($1\,\sigma$ uncertainties of $-0.37$ and $+0.59$), which we propagated to estimate the $1\,\sigma$ uncertainties on median $\mathcal{R}$. A robust linear regression obtained with the BCES $y|x$ method yielded a best-fitting slope of $\beta = 0.39 \pm 0.13$ between $\mathcal{R}$ and $\log~\Sigma$ (see Fig.~\ref{fig:r-sigm-m}). Additionally, we have calculated the Pearson correlation coefficient between $\mathcal{R}$ and $\log~\Sigma$ to be $r=0.20$ with a $3.3\,\sigma$ significance, indicating a statistically significant, albeit weak, positive correlation between $\mathcal{R}$ and $\log~\Sigma$. 

To test the robustness of our results, we evaluated median $\mathcal{R}$ across low, intermediate and high-density local environments using alternative $\log\,\Sigma$ cuts (see Table~\ref{tab:tab2}). While smaller subsample sizes (cuts 4 and 5) and reduced contrast between subsamples (cuts 3 and 6) naturally decreased the Anderson--Darling test statistical significance, the positive trends between $\mathcal{R}$ and $\log\,\Sigma$ remained consistent with the trend obtained using the fiducial cuts at the 33.3rd and 66.7th percentiles of $\log\,\Sigma$. 

The Pearson correlation, alongside the outcome of the Anderson--Darling test and the BCES $y|x$ linear regression, is consistent with the scenario in which the rotation rates of bars are influenced by the mechanism behind their formation, with denser environments favouring the formation of slower, tidal bars. This is in support of the predictions from simulations by \citet{1998ApJ...499..149M}, \citet{2004MNRAS.347..220B}, \citet{2014MNRAS.445.1339L,2016ApJ...826..227L},
\citet{2017MNRAS.464.1502M}, \citet{2017ApJ...842...56G} and \citet{2018ApJ...857....6L}. It is however important to note that these trends may also reflect variations in bar age or bar slowdown timescales, rather than a direct environmental effect. \citet{2025ApJ...979...60Z} argue that, when bar strength is taken into account, only galaxies which would not otherwise be able to form bars in isolation develop tidal bars which are slower than internally triggered bars. They propose that the differences in $\mathcal{R}$ are driven by the internal properties of the host galaxies, rather than the bar formation mechanism. 

The Tremaine--Weinberg method can only be applied to galaxies which meet strict criteria. Most importantly, the TW method assumes continuity of the tracer \citep{1984ApJ...282L...5T}. This condition will not be met in a galaxy with a disturbed velocity field due to environmental mechanisms (tidal interactions, ram-pressure stripping, etc.). As a result, galaxies in high-density environments often do not meet the criteria for TW and are discarded in bar studies. This may introduce bias against highly disturbed cluster galaxies and may reduce the sensitivity to the effects of extremely dense environments on bar rotation rates. 

As an additional check, we verified that $\Omega _{\mathrm{bar}}$, $R_{\mathrm{cr}}$ and $R_{\mathrm{bar}}$ show no significant correlations with $\log\,\Sigma$ in the \citet{2023MNRAS.521.1775G} sample.   

\subsubsection{The Milky Way bar}
Although we did not include the Milky Way (MW) in our statistical analysis, it is instructive to consider the MW's bar in the context of the observed relationship between $\mathcal{R}$ and $\log\,\Sigma$. Following the definition of $\log\,\Sigma$ from \citet{2006MNRAS.373..469B}, we approximated the MW's local environmental density in Appendix~\ref{apA} as $\log\,\Sigma=-0.94$, placing the MW in a low-density environment. Studies typically place the MW's bar rotation rate at $\mathcal{R} \sim 1.2$, classifying it as a fast rotator (\citealt{2017MNRAS.465.1621P}; \citealt*{2020RAA....20..159S}). Therefore, the MW's bar rotation rate is qualitatively consistent with the observed tendency for low-density environments to preferentially host faster bars. We note however that $\mathcal{R}$ evolves over time, so the MW's current bar rotation rate represents only its present evolutionary state, and the above consistency may be coincidental. None the less, the above result suggests that although the MW is atypical in colour-magnitude space (\citealt*{2011ApJ...736...84M}), its bar is not atypical compared to the galaxies in our sample.

\subsection{Ultrafast bars $\mathcal{R} < 1$ do not drive the observed trends} \label{uf}
In theory, bars which extend beyond $R_{\rm cr}$ should be unstable and rapidly dissolve (\citealt{1980A&A....81..198C,1992MNRAS.259..345A}). Despite this, bars with $\mathcal{R}<1$ are repeatedly measured observationally (\citealt{2008MNRAS.388.1803R,2017ApJ...835..279F,2019MNRAS.482.1733G}). Although these have been suggested to appear due to overestimated values of $R_{\mathrm{bar}}$ (\citealt{2021MNRAS.503.2833R,2021A&A...649A..30C}), the existence of ultrafast bars is widely debated (\citealt{2015A&A...576A.102A,2019MNRAS.482.1733G,2020A&A...641A.111C,2023MNRAS.521.1775G}), and we choose not to exclude them from our analysis. The issue requires further investigation and we do not resolve it this paper. 

We do however test the robustness of our results to the exclusion of bars with $\mathcal{R}<1$. Interestingly, we have found that the fraction of bars with $\mathcal{R}<1$ in the subgroups with $\log~\Sigma$ cut-offs at $-0.46$ and $0.13$ decreases with increasing environmental density ($19$, $16$, and $11$ per cent from low to high-density local environments), which is consistent with the observed trend that bars in denser environments tend to have higher median $\mathcal{R}$. Excluding ultrafast bars reduced the significance of the Anderson--Darling test between $\mathcal{R}$ distributions in low and high-density environments from $3.1\,\sigma$ to $2.7\,\sigma$. Although this falls slightly below $3\,\sigma$, the trend across the low, intermediate, and high-density subgroups remained consistent (median $\mathcal{R}$ of $1.54^{+0.10}_{-0.10}$, $1.70^{+0.12}_{-0.11}$ and $1.78^{+0.15}_{-0.13}$, respectively), with median $\mathcal{R}$ increasing by $\sim 10$ per cent in  each environment. Adopting a more conservative definition of ultrafast bars, as those with the upper $1\,\sigma$ limit below $\mathcal{R}=1$ (\citealt{2019MNRAS.482.1733G,2023MNRAS.521.1775G}), we identified 4, 3 and 2 ultrafast bars in the low, intermediate and high-density environments, respectively. Excluding the ultrafast bars resulted in  median $\mathcal{R}$ in the subgroups equal to $\mathcal{R} = 1.43^{+0.09}_{-0.08}$, $\mathcal{R} = 1.56^{+0.12}_{-0.10}$ and $\mathcal{R} = 1.67^{+0.13}_{-0.11}$, respectively, and a statistically significant Anderson--Darling test ($p_{\mathrm{AD}}=0.003$; $3.0\,\sigma$). The above results suggest that even if the systems which appear to have ultrafast bars are excluded from the analysis, the positive correlation remains. In fact, removing these bars disproportionately reduced the sample size in the low-density bin, which could itself be the source of the decreased Anderson--Darling test significance. While we cannot rule out the possibility that bars with $\mathcal{R}<1$ arise due to overestimated values of $R_{\mathrm{bar}}$ (\citealt{2021MNRAS.503.2833R,2021A&A...649A..30C}), their distribution across local density bins is consistent with the observed positive trend between $\mathcal{R}$ and $\log\,\Sigma$.

\subsection{Galaxy stellar mass $ \log M_\star $}
To test whether galaxy stellar mass could be responsible for variations in $\mathcal{R}$, we compared the distributions of $\mathcal{R}$ across low, intermediate and high-mass barred galaxies, with their median values of $\mathcal{R}$ being $\mathcal{R}~=~1.49_{-0.07}^{+0.07}$, $\mathcal{R}~=~1.58_{-0.06}^{+0.06}$ and $\mathcal{R}~=~1.40_{-0.07}^{+0.08}$, respectively. The Anderson--Darling test revealed no significant differences between the distributions of $\mathcal{R}$ in the low and high-mass subgroups ($p_{\mathrm{AD}}= 0.17$, $1.4\,\sigma$; see Fig.~\ref{fig:r-sigm-m}). It is however important to note the small sample size in this study. This issue could be addressed by increasing the sample size of barred galaxies for which galaxy stellar mass is known and focusing specifically on the relationship between galaxy stellar mass and bar rotation rate. The upcoming Integral Field Unit survey conducted by Hector \citep{2020SPIE11447E..15B} is of particular interest, as it aims to survey approximately 15,000 galaxies. Assuming that due to the constraints on the TW method only 2 per cent of the data are retained, this would allow to add roughly 300 galaxies to the sample.

\section{CONCLUSIONS} \label{con}

We have used the estimates of $\mathcal{R}$ obtained with the Tremaine--Weinberg method applied to IFU spectroscopy from SDSS MaNGA and CALIFA across six studies (\citealt{2015AA...576A.102A,2019MNRAS.482.1733G,2019AA...632A..51C,2020MNRAS.491.3655G,2022MNRAS.517.5660G,2023MNRAS.521.1775G}). Our sample consisted of 334 galaxies in the local Universe ($0.017 < z < 0.078$), which we cross-matched with catalogues containing environmental densities \citep{2006MNRAS.373..469B} and galactic stellar masses (\citealt{2003MNRAS.341...33K}), marking the first empirical test of the assumption that fast bars are formed by global instabilities in the galaxy disc, while slow bars are triggered or influenced by tidal interactions. 

We have found significant evidence consistent with the scenario in which bar rotation rates correlate with formation history, with denser environments preferentially hosting slower bars consistent with tidal formation ($p_{\mathrm{AD}}= 0.002$, $3.1\,\,\sigma$). Specifically, we have measured the median $\mathcal{R}$ in low ($\log \Sigma < -0.46$) and high-density ($\log \Sigma > 0.13$) environments to be $\mathcal{R} =1.39^{+0.09}_{-0.08}$ and $\mathcal{R} = 1.65^{+0.13}_{-0.11}$, respectively. Although the correlation between $\mathcal{R}$ and $\log\,\Sigma$ was statistically significant, its strength was modest ($r=0.2$; $3.3\,\sigma$), indicating a subtle environmental dependence. We subsequently tested the robustness of our results against different density cut-offs. Although all subdivisions led to reduced Anderson--Darling test significance, the positive trend between $\mathcal{R}$ and $\log\,\Sigma$ remained consistent with our hypothesis. Additionally, we have found no evidence suggesting significant differences in the bar pattern speed between low and high-mass galaxies. 

These results will be solidified by new Integral Field Unit data. An exciting upcoming survey will be conducted by Hector, a spectrograph installed on the Anglo-Australian Telescope. Hector aims to survey approximately 15,000 galaxies up to 2 effective radii \citep{2020SPIE11447E..15B}. This extensive survey will significantly increase the number of galaxies to which the Tremaine--Weinberg method can be applied, adding roughly 300 galaxies to the sample in this paper. We encourage observations of barred galaxies with IFUs as a means of providing insight into their formation and evolution. 

\section*{Acknowledgements}
We thank the referee for the interesting idea to investigate whether the Milky Way bar follows the observed relationship between $\mathcal{R}$ and environment. We also thank Yirui Zheng for the helpful discussion on tidal bars, and Marcin Semczuk for suggesting that we comment on the correlation between $\log\,\Sigma$ and $\Omega _{\mathrm{bar}}$, $R_{\mathrm{cr}}$ and $R_{\mathrm{bar}}$.  

\section*{Data Availability}   
This research has made use of publicly available data. The $\mathcal{R}$ data were obtained from six independent sources: \citet{2023MNRAS.521.1775G} (\url{https://zenodo.org/records/7567945}), \citet{2022MNRAS.517.5660G} (\url{https://github.com/lgarma/MWA_pattern_speed}), \citet[Table~4]{2020MNRAS.491.3655G}, \citet[Table~4]{2019MNRAS.482.1733G} (values taken from column $\mathcal{R}_\mathrm{l}$), \citet[Table~2]{2019AA...632A..51C} and \citet[Table~4]{2015AA...576A.102A} (values taken from column $\mathcal{R}_1$). Measurements of stellar masses ($\log M_\star$) and neighbour density ($\log~\Sigma$) were taken from publicly available catalogues cited in the text. The Catalog \& Atlas of the LV galaxies (\citealt{2013AJ....145..101K}) was used to obtain an estimate of $\log\,\Sigma$ for the Milky Way (Appendix~\ref{apA}).

\bibliographystyle{mnras}
\bibliography{bibliography}

@ARTICLE{2012MNRAS.424.2180M,
       author = {{Masters}, Karen L. and {Nichol}, Robert C. and {Haynes}, Martha P. and {Keel}, William C. and {Lintott}, Chris and {Simmons}, Brooke and {Skibba}, Ramin and {Bamford}, Steven and {Giovanelli}, Riccardo and {Schawinski}, Kevin},
        title = "{Galaxy Zoo and ALFALFA: atomic gas and the regulation of star formation in barred disc galaxies}",
      journal = {\mnras},
         year = 2012,
        month = aug,
       volume = {424},
       number = {3},
        pages = {2180-2192},
          doi = {10.1111/j.1365-2966.2012.21377.x},
archivePrefix = {arXiv},
       eprint = {1205.5271},
 primaryClass = {astro-ph.CO},
       adsurl = {https://ui.adsabs.harvard.edu/abs/2012MNRAS.424.2180M}
}

@ARTICLE{2013ApJ...779..162C,
       author = {{Cheung}, Edmond and {Athanassoula}, E. and {Masters}, Karen L. and {Nichol}, Robert C. and {Bosma}, A. and {Bell}, Eric F. and {Faber}, S.~M. and {Koo}, David C. and {Lintott}, Chris and {Melvin}, Thomas and {Schawinski}, Kevin and {Skibba}, Ramin A. and {Willett}, Kyle W.},
        title = "{Galaxy Zoo: Observing Secular Evolution through Bars}",
      journal = {\apj},
         year = 2013,
        month = dec,
       volume = {779},
       number = {2},
          eid = {162},
        pages = {162},
          doi = {10.1088/0004-637X/779/2/162},
archivePrefix = {arXiv},
       eprint = {1310.2941},
 primaryClass = {astro-ph.CO},
       adsurl = {https://ui.adsabs.harvard.edu/abs/2013ApJ...779..162C}
}

@ARTICLE{2021MNRAS.507.4389G,
       author = {{G{\'e}ron}, Tobias and {Smethurst}, R.~J. and {Lintott}, Chris and {Kruk}, Sandor and {Masters}, Karen L. and {Simmons}, Brooke and {Stark}, David V.},
        title = "{Galaxy zoo: stronger bars facilitate quenching in star-forming galaxies}",
      journal = {\mnras},
         year = 2021,
        month = nov,
       volume = {507},
       number = {3},
        pages = {4389-4408},
          doi = {10.1093/mnras/stab2064},
archivePrefix = {arXiv},
       eprint = {2107.06913},
 primaryClass = {astro-ph.GA},
       adsurl = {https://ui.adsabs.harvard.edu/abs/2021MNRAS.507.4389G}
}

@ARTICLE{2022MNRAS.517.5660G,
       author = {{Garma-Oehmichen}, L. and {Hern{\'a}ndez-Toledo}, H. and {Aquino-Ort{\'\i}z}, E. and {Martinez-Medina}, L. and {Puerari}, I. and {Cano-D{\'\i}az}, M. and {Valenzuela}, O. and {V{\'a}zquez-Mata}, J.~A. and {G{\'e}ron}, T. and {Mart{\'\i}nez-V{\'a}zquez}, L.~A. and {Lane}, R.},
        title = "{SDSS IV MaNGA: bar pattern speed in Milky Way analogue galaxies}",
      journal = {\mnras},
         year = 2022,
        month = dec,
       volume = {517},
       number = {4},
        pages = {5660-5677},
          doi = {10.1093/mnras/stac3069},
archivePrefix = {arXiv},
       eprint = {2210.11424},
 primaryClass = {astro-ph.GA},
       adsurl = {https://ui.adsabs.harvard.edu/abs/2022MNRAS.517.5660G}
}

@ARTICLE{2018MNRAS.473.4731K,
       author = {{Kruk}, Sandor J. and {Lintott}, Chris J. and {Bamford}, Steven P. and {Masters}, Karen L. and {Simmons}, Brooke D. and {H{\"a}u{\ss}ler}, Boris and {Cardamone}, Carolin N. and {Hart}, Ross E. and {Kelvin}, Lee and {Schawinski}, Kevin and {Smethurst}, Rebecca J. and {Vika}, Marina},
        title = "{Galaxy Zoo: secular evolution of barred galaxies from structural decomposition of multiband images}",
      journal = {\mnras},
         year = 2018,
        month = feb,
       volume = {473},
       number = {4},
        pages = {4731-4753},
          doi = {10.1093/mnras/stx2605},
archivePrefix = {arXiv},
       eprint = {1710.00093},
 primaryClass = {astro-ph.GA},
       adsurl = {https://ui.adsabs.harvard.edu/abs/2018MNRAS.473.4731K}
}

@ARTICLE{2014RvMP...86....1S,
       author = {{Sellwood}, J.~A.},
        title = "{Secular evolution in disk galaxies}",
      journal = {Rev. Mod. Phys.},
         year = 2014,
        month = jan,
       volume = {86},
       number = {1},
        pages = {1-46},
          doi = {10.1103/RevModPhys.86.1},
archivePrefix = {arXiv},
       eprint = {1310.0403},
 primaryClass = {astro-ph.GA},
       adsurl = {https://ui.adsabs.harvard.edu/abs/2014RvMP...86....1S}
}

@ARTICLE{2017MNRAS.464.1502M,
       author = {{Martinez-Valpuesta}, Inma and {Aguerri}, J. Alfonso L. and {Gonz{\'a}lez-Garc{\'\i}a}, A. C{\'e}sar and {Dalla Vecchia}, Claudio and {Stringer}, Martin},
        title = "{A numerical study of interactions and stellar bars}",
      journal = {\mnras},
         year = 2017,
        month = jan,
       volume = {464},
       number = {2},
        pages = {1502-1511},
          doi = {10.1093/mnras/stw2500},
archivePrefix = {arXiv},
       eprint = {1610.02326},
 primaryClass = {astro-ph.GA},
       adsurl = {https://ui.adsabs.harvard.edu/abs/2017MNRAS.464.1502M}
}

@ARTICLE{2017ApJ...842...56G,
       author = {{Gajda}, Grzegorz and {{\L}okas}, Ewa L. and {Athanassoula}, E.},
        title = "{Tidally Induced Bars in Dwarf Galaxies on Different Orbits around a Milky Way-like Host}",
      journal = {\apj},
         year = 2017,
        month = jun,
       volume = {842},
       number = {1},
          eid = {56},
        pages = {56},
          doi = {10.3847/1538-4357/aa74b4},
archivePrefix = {arXiv},
       eprint = {1703.02933},
 primaryClass = {astro-ph.GA},
       adsurl = {https://ui.adsabs.harvard.edu/abs/2017ApJ...842...56G}
}

@ARTICLE{2018ApJ...857....6L,
       author = {{{\L}okas}, Ewa L.},
        title = "{Formation of Tidally Induced Bars in Galactic Flybys: Prograde versus Retrograde Encounters}",
      journal = {\apj},
         year = 2018,
        month = apr,
       volume = {857},
       number = {1},
          eid = {6},
        pages = {6},
          doi = {10.3847/1538-4357/aab4ff},
archivePrefix = {arXiv},
       eprint = {1803.09465},
 primaryClass = {astro-ph.GA},
       adsurl = {https://ui.adsabs.harvard.edu/abs/2018ApJ...857....6L}
}

@ARTICLE{1984ApJ...282L...5T,
       author = {{Tremaine}, S. and {Weinberg}, M.~D.},
        title = "{A kinematic method for measuring the pattern speed of barred galaxies.}",
      journal = {\apjl},
         year = 1984,
        month = jul,
       volume = {282},
        pages = {L5-L7},
          doi = {10.1086/184292},
       adsurl = {https://ui.adsabs.harvard.edu/abs/1984ApJ...282L...5T}
}

@ARTICLE{2023MNRAS.521.1775G,
       author = {{G{\'e}ron}, Tobias and {Smethurst}, Rebecca J. and {Lintott}, Chris and {Kruk}, Sandor and {Masters}, Karen L. and {Simmons}, Brooke and {Mantha}, Kameswara Bharadwaj and {Walmsley}, Mike and {Garma-Oehmichen}, L. and {Drory}, Niv and {Lane}, Richard R.},
        title = "{Galaxy Zoo: kinematics of strongly and weakly barred galaxies}",
      journal = {\mnras},
         year = 2023,
        month = may,
       volume = {521},
       number = {2},
        pages = {1775-1793},
          doi = {10.1093/mnras/stad501},
archivePrefix = {arXiv},
       eprint = {2302.05464},
 primaryClass = {astro-ph.GA},
       adsurl = {https://ui.adsabs.harvard.edu/abs/2023MNRAS.521.1775G}
}

@ARTICLE{1987MNRAS.228..635N,
       author = {{Noguchi}, Masafumi},
        title = "{Close encounter between galaxies - II. Tidal deformation of a disc galaxy stabilized by massive halo.}",
      journal = {\mnras},
         year = 1987,
        month = oct,
       volume = {228},
        pages = {635-651},
          doi = {10.1093/mnras/228.3.635},
       adsurl = {https://ui.adsabs.harvard.edu/abs/1987MNRAS.228..635N}
}

@ARTICLE{1971ApJ...168..343H,
       author = {{Hohl}, Frank},
        title = "{Numerical Experiments with a Disk of Stars}",
      journal = {\apj},
         year = 1971,
        month = sep,
       volume = {168},
        pages = {343},
          doi = {10.1086/151091},
       adsurl = {https://ui.adsabs.harvard.edu/abs/1971ApJ...168..343H}
}

@ARTICLE{1973ApJ...186..467O,
       author = {{Ostriker}, J.~P. and {Peebles}, P.~J.~E.},
        title = "{A Numerical Study of the Stability of Flattened Galaxies: or, can Cold Galaxies Survive?}",
      journal = {\apj},
         year = 1973,
        month = dec,
       volume = {186},
        pages = {467-480},
          doi = {10.1086/152513},
       adsurl = {https://ui.adsabs.harvard.edu/abs/1973ApJ...186..467O}
}

@ARTICLE{1981A&A....96..164C,
       author = {{Combes}, F. and {Sanders}, R.~H.},
        title = "{Formation and properties of persisting stellar bars.}",
      journal = {\aap},
         year = 1981,
        month = mar,
       volume = {96},
        pages = {164-173},
       adsurl = {https://ui.adsabs.harvard.edu/abs/1981A&A....96..164C}
}

@ARTICLE{1991Natur.352..411R,
       author = {{Raha}, N. and {Sellwood}, J.~A. and {James}, R.~A. and {Kahn}, F.~D.},
        title = "{A dynamical instability of bars in disk galaxies}",
      journal = {\nat},
         year = 1991,
        month = aug,
       volume = {352},
       number = {6334},
        pages = {411-412},
          doi = {10.1038/352411a0},
       adsurl = {https://ui.adsabs.harvard.edu/abs/1991Natur.352..411R}
}

@ARTICLE{2017A&A...604A..75M,
       author = {{Moetazedian}, R. and {Polyachenko}, E.~V. and {Berczik}, P. and {Just}, A.},
        title = "{Effects of galaxy-satellite interactions on bar formation}",
      journal = {\aap},
         year = 2017,
        month = aug,
       volume = {604},
          eid = {A75},
        pages = {A75},
          doi = {10.1051/0004-6361/201630024},
archivePrefix = {arXiv},
       eprint = {1703.06002},
 primaryClass = {astro-ph.GA},
       adsurl = {https://ui.adsabs.harvard.edu/abs/2017A&A...604A..75M}
}

@ARTICLE{1998ApJ...499..149M,
       author = {{Miwa}, Toshinobu and {Noguchi}, Masafumi},
        title = "{Dynamical Properties of Tidally Induced Galactic Bars}",
      journal = {\apj},
         year = 1998,
        month = may,
       volume = {499},
       number = {1},
        pages = {149-166},
          doi = {10.1086/305611},
       adsurl = {https://ui.adsabs.harvard.edu/abs/1998ApJ...499..149M}
}

@ARTICLE{2016MNRAS.463.2210K,
       author = {{Kyziropoulos}, P.~E. and {Efthymiopoulos}, C. and {Gravvanis}, G.~A. and {Patsis}, P.~A.},
        title = "{Structures induced by companions in galactic discs}",
      journal = {\mnras},
         year = 2016,
        month = dec,
       volume = {463},
       number = {2},
        pages = {2210-2228},
          doi = {10.1093/mnras/stw2084},
archivePrefix = {arXiv},
       eprint = {1611.04891},
 primaryClass = {astro-ph.GA},
       adsurl = {https://ui.adsabs.harvard.edu/abs/2016MNRAS.463.2210K}
}

@ARTICLE{2017ApJ...835..279F,
       author = {{Font}, J. and {Beckman}, J.~E. and {Mart{\'\i}nez-Valpuesta}, I. and {Borlaff}, A.~S. and {James}, P.~A. and {D{\'\i}az-Garc{\'\i}a}, S. and {Garc{\'\i}a-Lorenzo}, B. and {Camps-Fari{\~n}a}, A. and {Guti{\'e}rrez}, L. and {Amram}, P.},
        title = "{Kinematic Clues to Bar Evolution for Galaxies in the Local Universe: Why the Fastest Rotating Bars are Rotating Most Slowly}",
      journal = {\apj},
         year = 2017,
        month = feb,
       volume = {835},
       number = {2},
          eid = {279},
        pages = {279},
          doi = {10.3847/1538-4357/835/2/279},
archivePrefix = {arXiv},
       eprint = {1702.01743},
 primaryClass = {astro-ph.GA},
       adsurl = {https://ui.adsabs.harvard.edu/abs/2017ApJ...835..279F}
}

@ARTICLE{2006MNRAS.373..469B,
       author = {{Baldry}, I.~K. and {Balogh}, M.~L. and {Bower}, R.~G. and {Glazebrook}, K. and {Nichol}, R.~C. and {Bamford}, S.~P. and {Budavari}, T.},
        title = "{Galaxy bimodality versus stellar mass and environment}",
      journal = {\mnras},
         year = 2006,
        month = dec,
       volume = {373},
       number = {2},
        pages = {469-483},
          doi = {10.1111/j.1365-2966.2006.11081.x},
archivePrefix = {arXiv},
       eprint = {astro-ph/0607648},
 primaryClass = {astro-ph},
       adsurl = {https://ui.adsabs.harvard.edu/abs/2006MNRAS.373..469B}
}

@ARTICLE{2009ApJS..182..543A,
       author = {{Abazajian}, Kevork N. and {Adelman-McCarthy}, Jennifer K. and {Ag{\"u}eros}, Marcel A. and {Allam}, Sahar S. and {Allende Prieto}, Carlos and {An}, Deokkeun and {Anderson}, Kurt S.~J. and {Anderson}, Scott F. and {Annis}, James and {Bahcall}, Neta A. and {Bailer-Jones}, C.~A.~L. and {Barentine}, J.~C. and {Bassett}, Bruce A. and {Becker}, Andrew C. and {Beers}, Timothy C. and {Bell}, Eric F. and {Belokurov}, Vasily and {Berlind}, Andreas A. and {Berman}, Eileen F. and {Bernardi}, Mariangela and {Bickerton}, Steven J. and {Bizyaev}, Dmitry and {Blakeslee}, John P. and {Blanton}, Michael R. and {Bochanski}, John J. and {Boroski}, William N. and {Brewington}, Howard J. and {Brinchmann}, Jarle and {Brinkmann}, J. and {Brunner}, Robert J. and {Budav{\'a}ri}, Tam{\'a}s and {Carey}, Larry N. and {Carliles}, Samuel and {Carr}, Michael A. and {Castander}, Francisco J. and {Cinabro}, David and {Connolly}, A.~J. and {Csabai}, Istv{\'a}n and {Cunha}, Carlos E. and {Czarapata}, Paul C. and {Davenport}, James R.~A. and {de Haas}, Ernst and {Dilday}, Ben and {Doi}, Mamoru and {Eisenstein}, Daniel J. and {Evans}, Michael L. and {Evans}, N.~W. and {Fan}, Xiaohui and {Friedman}, Scott D. and {Frieman}, Joshua A. and {Fukugita}, Masataka and {G{\"a}nsicke}, Boris T. and {Gates}, Evalyn and {Gillespie}, Bruce and {Gilmore}, G. and {Gonzalez}, Belinda and {Gonzalez}, Carlos F. and {Grebel}, Eva K. and {Gunn}, James E. and {Gy{\"o}ry}, Zsuzsanna and {Hall}, Patrick B. and {Harding}, Paul and {Harris}, Frederick H. and {Harvanek}, Michael and {Hawley}, Suzanne L. and {Hayes}, Jeffrey J.~E. and {Heckman}, Timothy M. and {Hendry}, John S. and {Hennessy}, Gregory S. and {Hindsley}, Robert B. and {Hoblitt}, J. and {Hogan}, Craig J. and {Hogg}, David W. and {Holtzman}, Jon A. and {Hyde}, Joseph B. and {Ichikawa}, Shin-ichi and {Ichikawa}, Takashi and {Im}, Myungshin and {Ivezi{\'c}}, {\v{Z}}eljko and {Jester}, Sebastian and {Jiang}, Linhua and {Johnson}, Jennifer A. and {Jorgensen}, Anders M. and {Juri{\'c}}, Mario and {Kent}, Stephen M. and {Kessler}, R. and {Kleinman}, S.~J. and {Knapp}, G.~R. and {Konishi}, Kohki and {Kron}, Richard G. and {Krzesinski}, Jurek and {Kuropatkin}, Nikolay and {Lampeitl}, Hubert and {Lebedeva}, Svetlana and {Lee}, Myung Gyoon and {Lee}, Young Sun and {French Leger}, R. and {L{\'e}pine}, S{\'e}bastien and {Li}, Nolan and {Lima}, Marcos and {Lin}, Huan and {Long}, Daniel C. and {Loomis}, Craig P. and {Loveday}, Jon and {Lupton}, Robert H. and {Magnier}, Eugene and {Malanushenko}, Olena and {Malanushenko}, Viktor and {Mandelbaum}, Rachel and {Margon}, Bruce and {Marriner}, John P. and {Mart{\'\i}nez-Delgado}, David and {Matsubara}, Takahiko and {McGehee}, Peregrine M. and {McKay}, Timothy A. and {Meiksin}, Avery and {Morrison}, Heather L. and {Mullally}, Fergal and {Munn}, Jeffrey A. and {Murphy}, Tara and {Nash}, Thomas and {Nebot}, Ada and {Neilsen}, Eric H., Jr. and {Newberg}, Heidi Jo and {Newman}, Peter R. and {Nichol}, Robert C. and {Nicinski}, Tom and {Nieto-Santisteban}, Maria and {Nitta}, Atsuko and {Okamura}, Sadanori and {Oravetz}, Daniel J. and {Ostriker}, Jeremiah P. and {Owen}, Russell and {Padmanabhan}, Nikhil and {Pan}, Kaike and {Park}, Changbom and {Pauls}, George and {Peoples}, John, Jr. and {Percival}, Will J. and {Pier}, Jeffrey R. and {Pope}, Adrian C. and {Pourbaix}, Dimitri and {Price}, Paul A. and {Purger}, Norbert and {Quinn}, Thomas and {Raddick}, M. Jordan and {Re Fiorentin}, Paola and {Richards}, Gordon T. and {Richmond}, Michael W. and {Riess}, Adam G. and {Rix}, Hans-Walter and {Rockosi}, Constance M. and {Sako}, Masao and {Schlegel}, David J. and {Schneider}, Donald P. and {Scholz}, Ralf-Dieter and {Schreiber}, Matthias R. and {Schwope}, Axel D. and {Seljak}, Uro{\v{s}} and {Sesar}, Branimir and {Sheldon}, Erin and {Shimasaku}, Kazu and {Sibley}, Valena C. and {Simmons}, A.~E. and {Sivarani}, Thirupathi and {Allyn Smith}, J. and {Smith}, Martin C. and {Smol{\v{c}}i{\'c}}, Vernesa and {Snedden}, Stephanie A. and {Stebbins}, Albert and {Steinmetz}, Matthias and {Stoughton}, Chris and {Strauss}, Michael A. and {SubbaRao}, Mark and {Suto}, Yasushi and {Szalay}, Alexander S. and {Szapudi}, Istv{\'a}n and {Szkody}, Paula and {Tanaka}, Masayuki and {Tegmark}, Max and {Teodoro}, Luis F.~A. and {Thakar}, Aniruddha R. and {Tremonti}, Christy A. and {Tucker}, Douglas L. and {Uomoto}, Alan and {Vanden Berk}, Daniel E. and {Vandenberg}, Jan and {Vidrih}, S. and {Vogeley}, Michael S. and {Voges}, Wolfgang and {Vogt}, Nicole P. and {Wadadekar}, Yogesh and {Watters}, Shannon and {Weinberg}, David H. and {West}, Andrew A. and {White}, Simon D.~M. and {Wilhite}, Brian C. and {Wonders}, Alainna C. and {Yanny}, Brian and {Yocum}, D.~R. and {York}, Donald G. and {Zehavi}, Idit and {Zibetti}, Stefano and {Zucker}, Daniel B.},
        title = "{The Seventh Data Release of the Sloan Digital Sky Survey}",
      journal = {\apjs},
         year = 2009,
        month = jun,
       volume = {182},
       number = {2},
        pages = {543-558},
          doi = {10.1088/0067-0049/182/2/543},
archivePrefix = {arXiv},
       eprint = {0812.0649},
 primaryClass = {astro-ph},
       adsurl = {https://ui.adsabs.harvard.edu/abs/2009ApJS..182..543A}
}

@ARTICLE{2022ApJS..259...35A,
       author = {{Abdurro'uf} and {Accetta}, Katherine and {Aerts}, Conny and {Silva Aguirre}, V{\'\i}ctor and {Ahumada}, Romina and {Ajgaonkar}, Nikhil and {Filiz Ak}, N. and {Alam}, Shadab and {Allende Prieto}, Carlos and {Almeida}, Andr{\'e}s and {Anders}, Friedrich and {Anderson}, Scott F. and {Andrews}, Brett H. and {Anguiano}, Borja and {Aquino-Ort{\'\i}z}, Erik and {Arag{\'o}n-Salamanca}, Alfonso and {Argudo-Fern{\'a}ndez}, Maria and {Ata}, Metin and {Aubert}, Marie and {Avila-Reese}, Vladimir and {Badenes}, Carles and {Barb{\'a}}, Rodolfo H. and {Barger}, Kat and {Barrera-Ballesteros}, Jorge K. and {Beaton}, Rachael L. and {Beers}, Timothy C. and {Belfiore}, Francesco and {Bender}, Chad F. and {Bernardi}, Mariangela and {Bershady}, Matthew A. and {Beutler}, Florian and {Bidin}, Christian Moni and {Bird}, Jonathan C. and {Bizyaev}, Dmitry and {Blanc}, Guillermo A. and {Blanton}, Michael R. and {Boardman}, Nicholas Fraser and {Bolton}, Adam S. and {Boquien}, M{\'e}d{\'e}ric and {Borissova}, Jura and {Bovy}, Jo and {Brandt}, W.~N. and {Brown}, Jordan and {Brownstein}, Joel R. and {Brusa}, Marcella and {Buchner}, Johannes and {Bundy}, Kevin and {Burchett}, Joseph N. and {Bureau}, Martin and {Burgasser}, Adam and {Cabang}, Tuesday K. and {Campbell}, Stephanie and {Cappellari}, Michele and {Carlberg}, Joleen K. and {Wanderley}, F{\'a}bio Carneiro and {Carrera}, Ricardo and {Cash}, Jennifer and {Chen}, Yan-Ping and {Chen}, Wei-Huai and {Cherinka}, Brian and {Chiappini}, Cristina and {Choi}, Peter Doohyun and {Chojnowski}, S. Drew and {Chung}, Haeun and {Clerc}, Nicolas and {Cohen}, Roger E. and {Comerford}, Julia M. and {Comparat}, Johan and {da Costa}, Luiz and {Covey}, Kevin and {Crane}, Jeffrey D. and {Cruz-Gonzalez}, Irene and {Culhane}, Connor and {Cunha}, Katia and {Dai}, Y. Sophia and {Damke}, Guillermo and {Darling}, Jeremy and {Davidson}, James W., Jr. and {Davies}, Roger and {Dawson}, Kyle and {De Lee}, Nathan and {Diamond-Stanic}, Aleksandar M. and {Cano-D{\'\i}az}, Mariana and {S{\'a}nchez}, Helena Dom{\'\i}nguez and {Donor}, John and {Duckworth}, Chris and {Dwelly}, Tom and {Eisenstein}, Daniel J. and {Elsworth}, Yvonne P. and {Emsellem}, Eric and {Eracleous}, Mike and {Escoffier}, Stephanie and {Fan}, Xiaohui and {Farr}, Emily and {Feng}, Shuai and {Fern{\'a}ndez-Trincado}, Jos{\'e} G. and {Feuillet}, Diane and {Filipp}, Andreas and {Fillingham}, Sean P. and {Frinchaboy}, Peter M. and {Fromenteau}, Sebastien and {Galbany}, Llu{\'\i}s and {Garc{\'\i}a}, Rafael A. and {Garc{\'\i}a-Hern{\'a}ndez}, D.~A. and {Ge}, Junqiang and {Geisler}, Doug and {Gelfand}, Joseph and {G{\'e}ron}, Tobias and {Gibson}, Benjamin J. and {Goddy}, Julian and {Godoy-Rivera}, Diego and {Grabowski}, Kathleen and {Green}, Paul J. and {Greener}, Michael and {Grier}, Catherine J. and {Griffith}, Emily and {Guo}, Hong and {Guy}, Julien and {Hadjara}, Massinissa and {Harding}, Paul and {Hasselquist}, Sten and {Hayes}, Christian R. and {Hearty}, Fred and {Hern{\'a}ndez}, Jes{\'u}s and {Hill}, Lewis and {Hogg}, David W. and {Holtzman}, Jon A. and {Horta}, Danny and {Hsieh}, Bau-Ching and {Hsu}, Chin-Hao and {Hsu}, Yun-Hsin and {Huber}, Daniel and {Huertas-Company}, Marc and {Hutchinson}, Brian and {Hwang}, Ho Seong and {Ibarra-Medel}, H{\'e}ctor J. and {Chitham}, Jacob Ider and {Ilha}, Gabriele S. and {Imig}, Julie and {Jaekle}, Will and {Jayasinghe}, Tharindu and {Ji}, Xihan and {Johnson}, Jennifer A. and {Jones}, Amy and {J{\"o}nsson}, Henrik and {Katkov}, Ivan and {Khalatyan}, Arman, Dr. and {Kinemuchi}, Karen and {Kisku}, Shobhit and {Knapen}, Johan H. and {Kneib}, Jean-Paul and {Kollmeier}, Juna A. and {Kong}, Miranda and {Kounkel}, Marina and {Kreckel}, Kathryn and {Krishnarao}, Dhanesh and {Lacerna}, Ivan and {Lane}, Richard R. and {Langgin}, Rachel and {Lavender}, Ramon and {Law}, David R. and {Lazarz}, Daniel and {Leung}, Henry W. and {Leung}, Ho-Hin and {Lewis}, Hannah M. and {Li}, Cheng and {Li}, Ran and {Lian}, Jianhui and {Liang}, Fu-Heng and {Lin}, Lihwai and {Lin}, Yen-Ting and {Lin}, Sicheng and {Lintott}, Chris and {Long}, Dan and {Longa-Pe{\~n}a}, Pen{\'e}lope and {L{\'o}pez-Cob{\'a}}, Carlos and {Lu}, Shengdong and {Lundgren}, Britt F. and {Luo}, Yuanze and {Mackereth}, J. Ted and {de la Macorra}, Axel and {Mahadevan}, Suvrath and {Majewski}, Steven R. and {Manchado}, Arturo and {Mandeville}, Travis and {Maraston}, Claudia and {Margalef-Bentabol}, Berta and {Masseron}, Thomas and {Masters}, Karen L. and {Mathur}, Savita and {McDermid}, Richard M. and {Mckay}, Myles and {Merloni}, Andrea and {Merrifield}, Michael and {Meszaros}, Szabolcs and {Miglio}, Andrea and {Di Mille}, Francesco and {Minniti}, Dante and {Minsley}, Rebecca and {Monachesi}, Antonela and {Moon}, Jeongin and {Mosser}, Benoit and {Mulchaey}, John and {Muna}, Demitri and {Mu{\~n}oz}, Ricardo R. and {Myers}, Adam D. and {Myers}, Natalie and {Nadathur}, Seshadri and {Nair}, Preethi and {Nandra}, Kirpal and {Neumann}, Justus and {Newman}, Jeffrey A. and {Nidever}, David L. and {Nikakhtar}, Farnik and {Nitschelm}, Christian and {O'Connell}, Julia E. and {Garma-Oehmichen}, Luis and {Luan Souza de Oliveira}, Gabriel and {Olney}, Richard and {Oravetz}, Daniel and {Ortigoza-Urdaneta}, Mario and {Osorio}, Yeisson and {Otter}, Justin and {Pace}, Zachary J. and {Padilla}, Nelson and {Pan}, Kaike and {Pan}, Hsi-An and {Parikh}, Taniya and {Parker}, James and {Peirani}, Sebastien and {Pe{\~n}a Ram{\'\i}rez}, Karla and {Penny}, Samantha and {Percival}, Will J. and {Perez-Fournon}, Ismael and {Pinsonneault}, Marc and {Poidevin}, Fr{\'e}d{\'e}rick and {Poovelil}, Vijith Jacob and {Price-Whelan}, Adrian M. and {B{\'a}rbara de Andrade Queiroz}, Anna and {Raddick}, M. Jordan and {Ray}, Amy and {Rembold}, Sandro Barboza and {Riddle}, Nicole and {Riffel}, Rogemar A. and {Riffel}, Rog{\'e}rio and {Rix}, Hans-Walter and {Robin}, Annie C. and {Rodr{\'\i}guez-Puebla}, Aldo and {Roman-Lopes}, Alexandre and {Rom{\'a}n-Z{\'u}{\~n}iga}, Carlos and {Rose}, Benjamin and {Ross}, Ashley J. and {Rossi}, Graziano and {Rubin}, Kate H.~R. and {Salvato}, Mara and {S{\'a}nchez}, Seb{\'a}stian F. and {S{\'a}nchez-Gallego}, Jos{\'e} R. and {Sanderson}, Robyn and {Santana Rojas}, Felipe Antonio and {Sarceno}, Edgar and {Sarmiento}, Regina and {Sayres}, Conor and {Sazonova}, Elizaveta and {Schaefer}, Adam L. and {Schiavon}, Ricardo and {Schlegel}, David J. and {Schneider}, Donald P. and {Schultheis}, Mathias and {Schwope}, Axel and {Serenelli}, Aldo and {Serna}, Javier and {Shao}, Zhengyi and {Shapiro}, Griffin and {Sharma}, Anubhav and {Shen}, Yue and {Shetrone}, Matthew and {Shu}, Yiping and {Simon}, Joshua D. and {Skrutskie}, M.~F. and {Smethurst}, Rebecca and {Smith}, Verne and {Sobeck}, Jennifer and {Spoo}, Taylor and {Sprague}, Dani and {Stark}, David V. and {Stassun}, Keivan G. and {Steinmetz}, Matthias and {Stello}, Dennis and {Stone-Martinez}, Alexander and {Storchi-Bergmann}, Thaisa and {Stringfellow}, Guy S. and {Stutz}, Amelia and {Su}, Yung-Chau and {Taghizadeh-Popp}, Manuchehr and {Talbot}, Michael S. and {Tayar}, Jamie and {Telles}, Eduardo and {Teske}, Johanna and {Thakar}, Ani and {Theissen}, Christopher and {Tkachenko}, Andrew and {Thomas}, Daniel and {Tojeiro}, Rita and {Hernandez Toledo}, Hector and {Troup}, Nicholas W. and {Trump}, Jonathan R. and {Trussler}, James and {Turner}, Jacqueline and {Tuttle}, Sarah and {Unda-Sanzana}, Eduardo and {V{\'a}zquez-Mata}, Jos{\'e} Antonio and {Valentini}, Marica and {Valenzuela}, Octavio and {Vargas-Gonz{\'a}lez}, Jaime and {Vargas-Maga{\~n}a}, Mariana and {Alfaro}, Pablo Vera and {Villanova}, Sandro and {Vincenzo}, Fiorenzo and {Wake}, David and {Warfield}, Jack T. and {Washington}, Jessica Diane and {Weaver}, Benjamin Alan and {Weijmans}, Anne-Marie and {Weinberg}, David H. and {Weiss}, Achim and {Westfall}, Kyle B. and {Wild}, Vivienne and {Wilde}, Matthew C. and {Wilson}, John C. and {Wilson}, Robert F. and {Wilson}, Mikayla and {Wolf}, Julien and {Wood-Vasey}, W.~M. and {Yan}, Renbin and {Zamora}, Olga and {Zasowski}, Gail and {Zhang}, Kai and {Zhao}, Cheng and {Zheng}, Zheng and {Zheng}, Zheng and {Zhu}, Kai},
        title = "{The Seventeenth Data Release of the Sloan Digital Sky Surveys: Complete Release of MaNGA, MaStar, and APOGEE-2 Data}",
      journal = {\apjs},
         year = 2022,
        month = apr,
       volume = {259},
       number = {2},
          eid = {35},
        pages = {35},
          doi = {10.3847/1538-4365/ac4414},
archivePrefix = {arXiv},
       eprint = {2112.02026},
 primaryClass = {astro-ph.GA},
       adsurl = {https://ui.adsabs.harvard.edu/abs/2022ApJS..259...35A}
}

@ARTICLE{2020MNRAS.491.3655G,
       author = {{Garma-Oehmichen}, L. and {Cano-D{\'\i}az}, M. and {Hern{\'a}ndez-Toledo}, H. and {Aquino-Ort{\'\i}z}, E. and {Valenzuela}, O. and {Aguerri}, J.~A.~L. and {S{\'a}nchez}, S.~F. and {Merrifield}, M.},
        title = "{SDSS-IV MaNGA: bar pattern speed estimates with the Tremaine-Weinberg method and their error sources}",
      journal = {\mnras},
         year = 2020,
        month = jan,
       volume = {491},
       number = {3},
        pages = {3655-3671},
          doi = {10.1093/mnras/stz3101},
archivePrefix = {arXiv},
       eprint = {1911.00090},
 primaryClass = {astro-ph.GA},
       adsurl = {https://ui.adsabs.harvard.edu/abs/2020MNRAS.491.3655G}
}

@ARTICLE{2019MNRAS.482.1733G,
       author = {{Guo}, Rui and {Mao}, Shude and {Athanassoula}, E. and {Li}, Hongyu and {Ge}, Junqiang and {Long}, R.~J. and {Merrifield}, Michael and {Masters}, Karen},
        title = "{SDSS-IV MaNGA: pattern speeds of barred galaxies}",
      journal = {\mnras},
         year = 2019,
        month = jan,
       volume = {482},
       number = {2},
        pages = {1733-1756},
          doi = {10.1093/mnras/sty2715},
archivePrefix = {arXiv},
       eprint = {1810.03257},
 primaryClass = {astro-ph.GA},
       adsurl = {https://ui.adsabs.harvard.edu/abs/2019MNRAS.482.1733G}
}

@ARTICLE{2015AA...576A.102A,
       author = {{Aguerri}, J.~A.~L. and {M{\'e}ndez-Abreu}, J. and {Falc{\'o}n-Barroso}, J. and {Amorin}, A. and {Barrera-Ballesteros}, J. and {Cid Fernandes}, R. and {Garc{\'\i}a-Benito}, R. and {Garc{\'\i}a-Lorenzo}, B. and {Gonz{\'a}lez Delgado}, R.~M. and {Husemann}, B. and {Kalinova}, V. and {Lyubenova}, M. and {Marino}, R.~A. and {M{\'a}rquez}, I. and {Mast}, D. and {P{\'e}rez}, E. and {S{\'a}nchez}, S.~F. and {van de Ven}, G. and {Walcher}, C.~J. and {Backsmann}, N. and {Cortijo-Ferrero}, C. and {Bland-Hawthorn}, J. and {del Olmo}, A. and {Iglesias-P{\'a}ramo}, J. and {P{\'e}rez}, I. and {S{\'a}nchez-Bl{\'a}zquez}, P. and {Wisotzki}, L. and {Ziegler}, B.},
        title = "{Bar pattern speeds in CALIFA galaxies. I. Fast bars across the Hubble sequence}",
      journal = {\aap},
         year = 2015,
        month = apr,
       volume = {576},
          eid = {A102},
        pages = {A102},
          doi = {10.1051/0004-6361/201423383},
archivePrefix = {arXiv},
       eprint = {1501.05498},
 primaryClass = {astro-ph.GA},
       adsurl = {https://ui.adsabs.harvard.edu/abs/2015A&A...576A.102A}
}

@ARTICLE{2008MNRAS.388.1803R,
       author = {{Rautiainen}, P. and {Salo}, H. and {Laurikainen}, E.},
        title = "{Model-based pattern speed estimates for 38 barred galaxies}",
      journal = {\mnras},
         year = 2008,
        month = aug,
       volume = {388},
       number = {4},
        pages = {1803-1818},
          doi = {10.1111/j.1365-2966.2008.13522.x},
archivePrefix = {arXiv},
       eprint = {0806.0471},
 primaryClass = {astro-ph},
       adsurl = {https://ui.adsabs.harvard.edu/abs/2008MNRAS.388.1803R}
}

@ARTICLE{2019AA...632A..51C,
       author = {{Cuomo}, Virginia and {Lopez Aguerri}, J. Alfonso and {Corsini}, Enrico Maria and {Debattista}, Victor P. and {M{\'e}ndez-Abreu}, Jairo and {Pizzella}, Alessandro},
        title = "{Bar pattern speeds in CALIFA galaxies. II. The case of weakly barred galaxies}",
      journal = {\aap},
         year = 2019,
        month = dec,
       volume = {632},
          eid = {A51},
        pages = {A51},
          doi = {10.1051/0004-6361/201936415},
archivePrefix = {arXiv},
       eprint = {1909.01023},
 primaryClass = {astro-ph.GA},
       adsurl = {https://ui.adsabs.harvard.edu/abs/2019A&A...632A..51C}
}

@ARTICLE{2023A&A...679A...5A,
       author = {{Aguerri}, J. Alfonso L. and {Cuomo}, Virginia and {Rojas-Roncero}, Azahara and {Morelli}, Lorenzo},
        title = "{Properties of barred galaxies with the environment. I. The case of the Virgo cluster}",
      journal = {\aap},
         year = 2023,
        month = nov,
       volume = {679},
          eid = {A5},
        pages = {A5},
          doi = {10.1051/0004-6361/202347500},
archivePrefix = {arXiv},
       eprint = {2309.11982},
 primaryClass = {astro-ph.GA},
       adsurl = {https://ui.adsabs.harvard.edu/abs/2023A&A...679A...5A}
}

@ARTICLE{2023MNRAS.526.4768W,
       author = {{Walmsley}, Mike and {G{\'e}ron}, Tobias and {Kruk}, Sandor and {Scaife}, Anna M.~M. and {Lintott}, Chris and {Masters}, Karen L. and {Dawson}, James M. and {Dickinson}, Hugh and {Fortson}, Lucy and {Garland}, Izzy L. and {Mantha}, Kameswara and {O'Ryan}, David and {Popp}, J{\"u}rgen and {Simmons}, Brooke and {Baeten}, Elisabeth M. and {Macmillan}, Christine},
        title = "{Galaxy Zoo DESI: Detailed morphology measurements for 8.7M galaxies in the DESI Legacy Imaging Surveys}",
      journal = {\mnras},
         year = 2023,
        month = dec,
       volume = {526},
       number = {3},
        pages = {4768-4786},
          doi = {10.1093/mnras/stad2919},
archivePrefix = {arXiv},
       eprint = {2309.11425},
 primaryClass = {astro-ph.GA},
       adsurl = {https://ui.adsabs.harvard.edu/abs/2023MNRAS.526.4768W}
}

@ARTICLE{2012ApJ...757...60K,
       author = {{Kraljic}, Katarina and {Bournaud}, Fr{\'e}d{\'e}ric and {Martig}, Marie},
        title = "{The Two-phase Formation History of Spiral Galaxies Traced by the Cosmic Evolution of the Bar Fraction}",
      journal = {\apj},
         year = 2012,
        month = sep,
       volume = {757},
       number = {1},
          eid = {60},
        pages = {60},
          doi = {10.1088/0004-637X/757/1/60},
archivePrefix = {arXiv},
       eprint = {1207.0351},
 primaryClass = {astro-ph.GA},
       adsurl = {https://ui.adsabs.harvard.edu/abs/2012ApJ...757...60K}
}

@ARTICLE{1980A&A....81..198C,
       author = {{Contopoulos}, G.},
        title = "{How far do bars extend}",
      journal = {\aap},
         year = 1980,
        month = jan,
       volume = {81},
       number = {1-2},
        pages = {198-209},
       adsurl = {https://ui.adsabs.harvard.edu/abs/1980A&A....81..198C}
}

@ARTICLE{1988A&A...189...42F,
       author = {{Fried}, J.~W.},
        title = "{How frequent are tidal interactions between galaxies ?}",
      journal = {\aap},
         year = 1988,
        month = jan,
       volume = {189},
        pages = {42-44},
       adsurl = {https://ui.adsabs.harvard.edu/abs/1988A&A...189...42F}
}

@INPROCEEDINGS{2020SPIE11447E..15B,
       author = {{Bryant}, Julia J. and {Bland-Hawthorn}, Joss and {Lawrence}, Jon and {Norris}, Barnaby and {Min}, Seong-Sik and {Brown}, Rebecca and {Wang}, Adeline and {Bhatia}, Gurashish Singh and {Saunders}, Will and {Content}, Robert and {Zhelem}, Ross and {Venkatesan}, Sudharshan and {Mohanan}, Mahesh and {Gillingham}, Peter and {Patterson}, Robert and {Robertson}, David and {Pai}, Naveen and {McGregor}, Helen and {Zheng}, Jessica and {Vaughan}, Sam and {Foster}, Caroline and {Leon-Saval}, Sergio and {Croom}, Scott},
        title = "{Hector: a new multi-object integral field spectrograph instrument for the Anglo-Australian Telescope}",
    booktitle = {Ground-based and Airborne Instrumentation for Astronomy VIII},
         year = 2020,
       editor = {{Evans}, Christopher J. and {Bryant}, Julia J. and {Motohara}, Kentaro},
       series = {Proc. SPIE Conf. Ser.},
    publisher = {SPIE},
       volume = {11447},
    address = {Bellingham},
        month = dec,
          eid = {1144715},
        pages = {1144715},
       adsurl = {https://ui.adsabs.harvard.edu/abs/2020SPIE11447E..15B}
}

@phdthesis{thesis,
  author  = {{G{\'e}ron}, Tobias},
  title   = "The impact of strong and weak bars on galaxy evolution",
  school  = "University of Oxford",
  year    = "2023"
}

@ARTICLE{2003MNRAS.341...33K,
       author = {{Kauffmann}, Guinevere and {Heckman}, Timothy M. and {White}, Simon D.~M. and {Charlot}, St{\'e}phane and {Tremonti}, Christy and {Brinchmann}, Jarle and {Bruzual}, Gustavo and {Peng}, Eric W. and {Seibert}, Mark and {Bernardi}, Mariangela and {Blanton}, Michael and {Brinkmann}, Jon and {Castander}, Francisco and {Cs{\'a}bai}, Istvan and {Fukugita}, Masataka and {Ivezic}, Zeljko and {Munn}, Jeffrey A. and {Nichol}, Robert C. and {Padmanabhan}, Nikhil and {Thakar}, Aniruddha R. and {Weinberg}, David H. and {York}, Donald},
        title = "{Stellar masses and star formation histories for {}10$^{5}$ galaxies from the Sloan Digital Sky Survey}",
      journal = {\mnras},
         year = 2003,
        month = may,
       volume = {341},
       number = {1},
        pages = {33-53},
          doi = {10.1046/j.1365-8711.2003.06291.x},
archivePrefix = {arXiv},
       eprint = {astro-ph/0204055},
 primaryClass = {astro-ph},
       adsurl = {https://ui.adsabs.harvard.edu/abs/2003MNRAS.341...33K}
}

@ARTICLE{2021MNRAS.501..994S,
       author = {{Sarkar}, Suman and {Pandey}, Biswajit and {Bhattacharjee}, Snehasish},
        title = "{Do galactic bars depend on environment?: an information theoretic analysis of Galaxy Zoo 2}",
      journal = {\mnras},
         year = 2021,
        month = feb,
       volume = {501},
       number = {1},
        pages = {994-1001},
          doi = {10.1093/mnras/staa3665},
archivePrefix = {arXiv},
       eprint = {2009.12797},
 primaryClass = {astro-ph.GA},
       adsurl = {https://ui.adsabs.harvard.edu/abs/2021MNRAS.501..994S}
}

@ARTICLE{2018MNRAS.477.1451F,
       author = {{Fujii}, M.~S. and {B{\'e}dorf}, J. and {Baba}, J. and {Portegies Zwart}, S.},
        title = "{The dynamics of stellar discs in live dark-matter haloes}",
      journal = {\mnras},
         year = 2018,
        month = jun,
       volume = {477},
       number = {2},
        pages = {1451-1471},
          doi = {10.1093/mnras/sty711},
archivePrefix = {arXiv},
       eprint = {1712.00058},
 primaryClass = {astro-ph.GA},
       adsurl = {https://ui.adsabs.harvard.edu/abs/2018MNRAS.477.1451F}
}

@ARTICLE{2023ApJ...947...80B,
       author = {{Bland-Hawthorn}, Joss and {Tepper-Garcia}, Thor and {Agertz}, Oscar and {Freeman}, Ken},
        title = "{The Rapid Onset of Stellar Bars in the Baryon-dominated Centers of Disk Galaxies}",
      journal = {\apj},
         year = 2023,
        month = apr,
       volume = {947},
       number = {2},
          eid = {80},
        pages = {80},
          doi = {10.3847/1538-4357/acc469},
archivePrefix = {arXiv},
       eprint = {2303.05574},
 primaryClass = {astro-ph.GA},
       adsurl = {https://ui.adsabs.harvard.edu/abs/2023ApJ...947...80B}
}

@ARTICLE{2020MNRAS.491.2547R,
       author = {{Rosas-Guevara}, Yetli and {Bonoli}, Silvia and {Dotti}, Massimo and {Zana}, Tommaso and {Nelson}, Dylan and {Pillepich}, Annalisa and {Ho}, Luis C. and {Izquierdo-Villalba}, David and {Hernquist}, Lars and {Pakmor}, R{\"u}ediger},
        title = "{The buildup of strongly barred galaxies in the TNG100 simulation}",
      journal = {\mnras},
         year = 2020,
        month = jan,
       volume = {491},
       number = {2},
        pages = {2547-2564},
          doi = {10.1093/mnras/stz3180},
archivePrefix = {arXiv},
       eprint = {1908.00547},
 primaryClass = {astro-ph.GA},
       adsurl = {https://ui.adsabs.harvard.edu/abs/2020MNRAS.491.2547R}
}

@ARTICLE{2022MNRAS.509.1262K,
       author = {{Kumar}, Ankit and {Das}, Mousumi and {Kataria}, Sandeep Kumar},
        title = "{The effect of dark matter halo shape on bar buckling and boxy/peanut bulges}",
      journal = {\mnras},
         year = 2022,
        month = jan,
       volume = {509},
       number = {1},
        pages = {1262-1268},
          doi = {10.1093/mnras/stab3019},
archivePrefix = {arXiv},
       eprint = {2110.08165},
 primaryClass = {astro-ph.GA},
       adsurl = {https://ui.adsabs.harvard.edu/abs/2022MNRAS.509.1262K}
}

@ARTICLE{2007MNRAS.382..801M,
       author = {{Mouhcine}, M. and {Baldry}, I.~K. and {Bamford}, S.~P.},
        title = "{The environmental dependence of the chemical properties of star-forming galaxies}",
      journal = {\mnras},
         year = 2007,
        month = dec,
       volume = {382},
       number = {2},
        pages = {801-808},
          doi = {10.1111/j.1365-2966.2007.12405.x},
archivePrefix = {arXiv},
       eprint = {0709.3794},
 primaryClass = {astro-ph},
       adsurl = {https://ui.adsabs.harvard.edu/abs/2007MNRAS.382..801M}
}

@ARTICLE{2010A&A...509A..40I,
       author = {{Iovino}, A. and {Cucciati}, O. and {Scodeggio}, M. and {Knobel}, C. and {Kova{\v{c}}}, K. and {Lilly}, S. and {Bolzonella}, M. and {Tasca}, L.~A.~M. and {Zamorani}, G. and {Zucca}, E. and {Caputi}, K. and {Pozzetti}, L. and {Oesch}, P. and {Lamareille}, F. and {Halliday}, C. and {Bardelli}, S. and {Finoguenov}, A. and {Guzzo}, L. and {Kampczyk}, P. and {Maier}, C. and {Tanaka}, M. and {Vergani}, D. and {Carollo}, C.~M. and {Contini}, T. and {Kneib}, J.-P. and {Le F{\`e}vre}, O. and {Mainieri}, V. and {Renzini}, A. and {Bongiorno}, A. and {Coppa}, G. and {de la Torre}, S. and {de Ravel}, L. and {Franzetti}, P. and {Garilli}, B. and {Le Borgne}, J.-F. and {Le Brun}, V. and {Mignoli}, M. and {Pell{\`o}}, R. and {Peng}, Y. and {Perez-Montero}, E. and {Ricciardelli}, E. and {Silverman}, J.~D. and {Tresse}, L. and {Abbas}, U. and {Bottini}, D. and {Cappi}, A. and {Cassata}, P. and {Cimatti}, A. and {Koekemoer}, A.~M. and {Leauthaud}, A. and {Maccagni}, D. and {Marinoni}, C. and {McCracken}, H.~J. and {Memeo}, P. and {Meneux}, B. and {Porciani}, C. and {Scaramella}, R. and {Schiminovich}, D. and {Scoville}, N.},
        title = "{The zCOSMOS redshift survey: how group environment alters global downsizing trends}",
      journal = {\aap},
         year = 2010,
        month = jan,
       volume = {509},
          eid = {A40},
        pages = {A40},
          doi = {10.1051/0004-6361/200912558},
archivePrefix = {arXiv},
       eprint = {0909.1951},
 primaryClass = {astro-ph.CO},
       adsurl = {https://ui.adsabs.harvard.edu/abs/2010A&A...509A..40I}
}

@ARTICLE{2009MNRAS.393.1324B,
       author = {{Bamford}, Steven P. and {Nichol}, Robert C. and {Baldry}, Ivan K. and {Land}, Kate and {Lintott}, Chris J. and {Schawinski}, Kevin and {Slosar}, An{\v{z}}e and {Szalay}, Alexander S. and {Thomas}, Daniel and {Torki}, Mehri and {Andreescu}, Dan and {Edmondson}, Edward M. and {Miller}, Christopher J. and {Murray}, Phil and {Raddick}, M. Jordan and {Vandenberg}, Jan},
        title = "{Galaxy Zoo: the dependence of morphology and colour on environment*}",
      journal = {\mnras},
         year = 2009,
        month = mar,
       volume = {393},
       number = {4},
        pages = {1324-1352},
          doi = {10.1111/j.1365-2966.2008.14252.x},
archivePrefix = {arXiv},
       eprint = {0805.2612},
 primaryClass = {astro-ph},
       adsurl = {https://ui.adsabs.harvard.edu/abs/2009MNRAS.393.1324B}
}

@ARTICLE{2000AJ....119..536E,
       author = {{Eskridge}, Paul B. and {Frogel}, Jay A. and {Pogge}, Richard W. and {Quillen}, Alice C. and {Davies}, Roger L. and {DePoy}, D.~L. and {Houdashelt}, Mark L. and {Kuchinski}, Leslie E. and {Ram{\'\i}rez}, Solange V. and {Sellgren}, K. and {Terndrup}, Donald M. and {Tiede}, Glenn P.},
        title = "{The Frequency of Barred Spiral Galaxies in the Near-Infrared}",
      journal = {\aj},
         year = 2000,
        month = feb,
       volume = {119},
       number = {2},
        pages = {536-544},
          doi = {10.1086/301203},
archivePrefix = {arXiv},
       eprint = {astro-ph/9910479},
 primaryClass = {astro-ph},
       adsurl = {https://ui.adsabs.harvard.edu/abs/2000AJ....119..536E}
}

@ARTICLE{2008ApJ...675.1141S,
       author = {{Sheth}, Kartik and {Elmegreen}, Debra Meloy and {Elmegreen}, Bruce G. and {Capak}, Peter and {Abraham}, Roberto G. and {Athanassoula}, E. and {Ellis}, Richard S. and {Mobasher}, Bahram and {Salvato}, Mara and {Schinnerer}, Eva and {Scoville}, Nicholas Z. and {Spalsbury}, Lori and {Strubbe}, Linda and {Carollo}, Marcella and {Rich}, Michael and {West}, Andrew A.},
        title = "{Evolution of the Bar Fraction in COSMOS: Quantifying the Assembly of the Hubble Sequence}",
      journal = {\apj},
         year = 2008,
        month = mar,
       volume = {675},
       number = {2},
        pages = {1141-1155},
          doi = {10.1086/524980},
archivePrefix = {arXiv},
       eprint = {0710.4552},
 primaryClass = {astro-ph},
       adsurl = {https://ui.adsabs.harvard.edu/abs/2008ApJ...675.1141S}
}

@ARTICLE{2018MNRAS.474.5372E,
       author = {{Erwin}, Peter},
        title = "{The dependence of bar frequency on galaxy mass, colour, and gas content - and angular resolution - in the local universe}",
      journal = {\mnras},
         year = 2018,
        month = mar,
       volume = {474},
       number = {4},
        pages = {5372-5392},
          doi = {10.1093/mnras/stx3117},
archivePrefix = {arXiv},
       eprint = {1711.04867},
 primaryClass = {astro-ph.GA},
       adsurl = {https://ui.adsabs.harvard.edu/abs/2018MNRAS.474.5372E}
}

@ARTICLE{2023Natur.623..499C,
       author = {{Costantin}, Luca and {P{\'e}rez-Gonz{\'a}lez}, Pablo G. and {Guo}, Yuchen and {Buttitta}, Chiara and {Jogee}, Shardha and {Bagley}, Micaela B. and {Barro}, Guillermo and {Kartaltepe}, Jeyhan S. and {Koekemoer}, Anton M. and {Cabello}, Cristina and {Corsini}, Enrico Maria and {M{\'e}ndez-Abreu}, Jairo and {de la Vega}, Alexander and {Iyer}, Kartheik G. and {Bisigello}, Laura and {Cheng}, Yingjie and {Morelli}, Lorenzo and {Arrabal Haro}, Pablo and {Buitrago}, Fernando and {Cooper}, M.~C. and {Dekel}, Avishai and {Dickinson}, Mark and {Finkelstein}, Steven L. and {Giavalisco}, Mauro and {Holwerda}, Benne W. and {Huertas-Company}, Marc and {Lucas}, Ray A. and {Papovich}, Casey and {Pirzkal}, Nor and {Seill{\'e}}, Lise-Marie and {Vega-Ferrero}, Jes{\'u}s and {Wuyts}, Stijn and {Yung}, L.~Y. Aaron},
        title = "{A Milky Way-like barred spiral galaxy at a redshift of 3}",
      journal = {\nat},
         year = 2023,
        month = nov,
       volume = {623},
       number = {7987},
        pages = {499-501},
          doi = {10.1038/s41586-023-06636-x},
archivePrefix = {arXiv},
       eprint = {2311.04283},
 primaryClass = {astro-ph.GA},
       adsurl = {https://ui.adsabs.harvard.edu/abs/2023Natur.623..499C}
}

@ARTICLE{2023ApJ...958...36S,
       author = {{Smail}, Ian and {Dudzevi{\v{c}}i{\={u}}t{\.{e}}}, Ugn{\.{e}} and {Gurwell}, Mark and {Fazio}, Giovanni G. and {Willner}, S.~P. and {Swinbank}, A.~M. and {Arumugam}, Vinodiran and {Summers}, Jake and {Cohen}, Seth H. and {Jansen}, Rolf A. and {Windhorst}, Rogier A. and {Meena}, Ashish and {Zitrin}, Adi and {Keel}, William C. and {Cheng}, Cheng and {Coe}, Dan and {Conselice}, Christopher J. and {D'Silva}, Jordan C.~J. and {Driver}, Simon P. and {Frye}, Brenda and {Grogin}, Norman A. and {Koekemoer}, Anton M. and {Marshall}, Madeline A. and {Nonino}, Mario and {Pirzkal}, Nor and {Robotham}, Aaron and {Rutkowski}, Michael J. and {Ryan}, Jr., Russell E. and {Tompkins}, Scott and {Willmer}, Christopher N.~A. and {Yan}, Haojing and {Broadhurst}, Thomas J. and {Diego}, Jos{\'e} M. and {Kamieneski}, Patrick and {Yun}, Min},
        title = "{Hidden Giants in JWST's PEARLS: An Ultramassive z = 4.26 Submillimeter Galaxy that Is Invisible to HST}",
      journal = {\apj},
         year = 2023,
        month = nov,
       volume = {958},
       number = {1},
          eid = {36},
        pages = {36},
          doi = {10.3847/1538-4357/acf931},
archivePrefix = {arXiv},
       eprint = {2306.16039},
 primaryClass = {astro-ph.GA},
       adsurl = {https://ui.adsabs.harvard.edu/abs/2023ApJ...958...36S}
}

@ARTICLE{2025MNRAS.537.1163A,
       author = {{Amvrosiadis}, A. and {Lange}, S. and {Nightingale}, J.~W. and {He}, Q. and {Frenk}, C.~S. and {Oman}, K.~A. and {Smail}, I. and {Swinbank}, A.~M. and {Fragkoudi}, F. and {Gadotti}, D.~A. and {Cole}, S. and {Borsato}, E. and {Robertson}, A. and {Massey}, R. and {Cao}, X. and {Li}, R.},
        title = "{The onset of bar formation in a massive galaxy at z \raisebox{-0.5ex}\textasciitilde 3.8}",
      journal = {\mnras},
         year = 2025,
        month = feb,
       volume = {537},
       number = {2},
        pages = {1163-1181},
          doi = {10.1093/mnras/staf048},
archivePrefix = {arXiv},
       eprint = {2404.01918},
 primaryClass = {astro-ph.GA},
       adsurl = {https://ui.adsabs.harvard.edu/abs/2025MNRAS.537.1163A}
}

@ARTICLE{2021MNRAS.502.3085G,
       author = {{Ghosh}, Soumavo and {Saha}, Kanak and {Di Matteo}, Paola and {Combes}, Francoise},
        title = "{Fate of stellar bars in minor merger of galaxies}",
      journal = {\mnras},
         year = 2021,
        month = apr,
       volume = {502},
       number = {2},
        pages = {3085-3100},
          doi = {10.1093/mnras/stab238},
archivePrefix = {arXiv},
       eprint = {2008.04942},
 primaryClass = {astro-ph.GA},
       adsurl = {https://ui.adsabs.harvard.edu/abs/2021MNRAS.502.3085G}
}

@ARTICLE{2020A&A...641A..77C,
       author = {{Cavanagh}, M.~K. and {Bekki}, K.},
        title = "{Bars formed in galaxy merging and their classification with deep learning}",
      journal = {\aap},
         year = 2020,
        month = sep,
       volume = {641},
          eid = {A77},
        pages = {A77},
          doi = {10.1051/0004-6361/202037963},
archivePrefix = {arXiv},
       eprint = {2006.14847},
 primaryClass = {astro-ph.GA},
       adsurl = {https://ui.adsabs.harvard.edu/abs/2020A&A...641A..77C}
}

@ARTICLE{2014MNRAS.438.2882M,
       author = {{Melvin}, Thomas and {Masters}, Karen and {Lintott}, Chris and {Nichol}, Robert C. and {Simmons}, Brooke and {Bamford}, Steven P. and {Casteels}, Kevin R.~V. and {Cheung}, Edmond and {Edmondson}, Edward M. and {Fortson}, Lucy and {Schawinski}, Kevin and {Skibba}, Ramin A. and {Smith}, Arfon M. and {Willett}, Kyle W.},
        title = "{Galaxy Zoo: an independent look at the evolution of the bar fraction over the last eight billion years from HST-COSMOS}",
      journal = {\mnras},
         year = 2014,
        month = mar,
       volume = {438},
       number = {4},
        pages = {2882-2897},
          doi = {10.1093/mnras/stt2397},
archivePrefix = {arXiv},
       eprint = {1401.3334},
 primaryClass = {astro-ph.GA},
       adsurl = {https://ui.adsabs.harvard.edu/abs/2014MNRAS.438.2882M}
}

@ARTICLE{2014ApJ...790L..33L,
       author = {{Lang}, Meagan and {Holley-Bockelmann}, Kelly and {Sinha}, Manodeep},
        title = "{Bar Formation from Galaxy Flybys}",
      journal = {\apjl},
         year = 2014,
        month = aug,
       volume = {790},
       number = {2},
          eid = {L33},
        pages = {L33},
          doi = {10.1088/2041-8205/790/2/L33},
archivePrefix = {arXiv},
       eprint = {1405.5832},
 primaryClass = {astro-ph.GA},
       adsurl = {https://ui.adsabs.harvard.edu/abs/2014ApJ...790L..33L}
}

@ARTICLE{2013MNRAS.429.1949A,
       author = {{Athanassoula}, E. and {Machado}, Rubens E.~G. and {Rodionov}, S.~A.},
        title = "{Bar formation and evolution in disc galaxies with gas and a triaxial halo: morphology, bar strength and halo properties}",
      journal = {\mnras},
         year = 2013,
        month = mar,
       volume = {429},
       number = {3},
        pages = {1949-1969},
          doi = {10.1093/mnras/sts452},
archivePrefix = {arXiv},
       eprint = {1211.6754},
 primaryClass = {astro-ph.CO},
       adsurl = {https://ui.adsabs.harvard.edu/abs/2013MNRAS.429.1949A}
}

@ARTICLE{2025ApJ...979...60Z,
       author = {{Zheng}, Yirui and {Shen}, Juntai},
        title = "{Comparison of Bar Formation Mechanisms. I. Does a Tidally Induced Bar Rotate Slower than an Internally Induced Bar?}",
      journal = {\apj},
         year = 2025,
        month = jan,
       volume = {979},
       number = {1},
          eid = {60},
        pages = {60},
          doi = {10.3847/1538-4357/ad9bae},
archivePrefix = {arXiv},
       eprint = {2412.04770},
 primaryClass = {astro-ph.GA},
       adsurl = {https://ui.adsabs.harvard.edu/abs/2025ApJ...979...60Z}
}

@ARTICLE{2004ApJ...613L..29M,
       author = {{Martinez-Valpuesta}, Inma and {Shlosman}, Isaac},
        title = "{Why Buckling Stellar Bars Weaken in Disk Galaxies}",
      journal = {\apjl},
         year = 2004,
        month = sep,
       volume = {613},
       number = {1},
        pages = {L29-L32},
          doi = {10.1086/424876},
archivePrefix = {arXiv},
       eprint = {astro-ph/0408241},
 primaryClass = {astro-ph},
       adsurl = {https://ui.adsabs.harvard.edu/abs/2004ApJ...613L..29M}
}

@ARTICLE{2006ApJ...637..214M,
       author = {{Martinez-Valpuesta}, Inma and {Shlosman}, Isaac and {Heller}, Clayton},
        title = "{Evolution of Stellar Bars in Live Axisymmetric Halos: Recurrent Buckling and Secular Growth}",
      journal = {\apj},
         year = 2006,
        month = jan,
       volume = {637},
       number = {1},
        pages = {214-226},
          doi = {10.1086/498338},
archivePrefix = {arXiv},
       eprint = {astro-ph/0507219},
 primaryClass = {astro-ph},
       adsurl = {https://ui.adsabs.harvard.edu/abs/2006ApJ...637..214M}
}

@incollection{2013seg..book..305A,
  author       = {Athanassoula, E.},
  year         = {2013},
  title        = {Bars and secular evolution in disk galaxies: Theoretical input},
  booktitle    = {Secular Evolution of Galaxies. Proceedings of the XXIII Canary Islands Winter School of Astrophysics},
  editor       = {Falc{\'o}n-Barroso, J. and Knapen, J. H.},
  publisher    = {Cambridge Univ. Press},
  address      = {Cambridge},
  pages        = {305},
}

\appendix

\section{derivation of $\log\,\Sigma$ for the Milky Way} \label{apA}

To place the Milky Way in the context of the observed relationship between bar rotation rate and environment, we estimate its local environmental density, following the definition of $\log\,\Sigma$ from \citet{2006MNRAS.373..469B}, outlined in Section \ref{log_Sigma}. This estimate is only approximate, since the MW cannot be treated identically to galaxies in redshift surveys. 

\citet{2006MNRAS.373..469B} identified the 4th and 5th nearest neighbours among galaxies with $r$-band absolute magnitudes $M_r<-20$ that are within $1000\,\rm km \, s^{-1}$ of the galaxy for which $\log\,\Sigma$ is being measured. As no complete $r$-band catalogue exists for the Local Volume, we instead identify the nearest significant neighbours among galaxies with extinction-corrected absolute $B$-band magnitudes $M_B^c <-19$.

We use the latest version of the Catalog \& Atlas of the LV galaxies (LVG; \citealt{2013AJ....145..101K}) which contains 1675 galaxies within $11\, \rm MPc$ of the MW with radial velocities with respect to the centroid of the Local Group within $600\,\rm km \, s^{-1}$. Unlike the projected distances used by \citet{2006MNRAS.373..469B}, we adopt three-dimensional distances to nearby galaxies available in the LVG catalogue.

We identify \mbox{NGC 4945} ($d_4 = 3.47\,\textrm{Mpc}$) and \mbox{Messier 82} (${d_5 = 3.61\textrm{Mpc}}$) as the 4th and 5th nearest neighbours of the MW, yielding $\log\,\Sigma_{4} = -0.98$ and $\log\,\Sigma_{5} = -0.91$. Following \citet{2006MNRAS.373..469B}, we approximate the MW's local density as $\log\,\Sigma = -0.94$, placing it in a low-density environment, consistent with typical definitions of voids (\citealt{2006MNRAS.373..469B,2007MNRAS.382..801M}). The LVG catalogue does not provide distance uncertainties, however, varying distance by $10-15$ per cent translates to a ${\lesssim 0.1\, \rm dex}$ shift in $\log\,\Sigma$, which does not change the classification of the MW's environment as low-density.

\bsp	

\label{lastpage}
\end{document}